\documentclass[lettersize,journal]{IEEEtran}
\IEEEoverridecommandlockouts
\usepackage{cite}
\usepackage{amsmath,amssymb,amsfonts}
\usepackage{graphicx}
\usepackage{textcomp}
\usepackage{xcolor}
\usepackage{cleveref}
\def\BibTeX{{\rm B\kern-.05em{\sc i\kern-.025em b}\kern-.08em
		T\kern-.1667em\lower.7ex\hbox{E}\kern-.125emX}}
\usepackage{multirow}
\usepackage{booktabs}
\usepackage{amsfonts}
\usepackage{amsmath}
\usepackage{amsmath,cases}
\usepackage{amssymb}
\usepackage{blindtext, graphicx}
\usepackage{cite}
\usepackage{epsfig}
\usepackage{url}
\usepackage{algorithm}
\usepackage{algorithmicx, algpseudocode}
\usepackage{epstopdf}
\usepackage{color}
\usepackage[tight]{subfigure}
\usepackage{mathrsfs}
\usepackage[justification=centering]{caption}
\usepackage{float}
\usepackage{cuted}
\usepackage{lipsum}
\usepackage{diagbox}
\usepackage{graphicx,tabularx}
\usepackage{lipsum}
\usepackage{enumitem}
\usepackage{academicons}
\usepackage{xcolor}

\newcommand{\orcid}[1]{\href{https://orcid.org/#1}{\textcolor[HTML]{A6CE39}{\aiOrcid}}}

\newcommand\blfootnote[1]{%
	\begingroup
	\renewcommand\thefootnote{}\footnote{#1}%
	\addtocounter{footnote}{-1}%
	\endgroup
}
\newtheorem{theorem}{\pmb{Theorem}}

\newcommand{\pmw}{\pmb{\text{w}}}
\newcommand{\pmW}{\pmb{\text{W}}}
\newcommand{\sigk}{\sigma_{k}}
\newcommand{\sige}{\sigma_{e}}
\newcommand{\btheta}{\pmb{\theta}}
\newcommand{\thetak}{\pmb{\theta}^{(\iota)}}
\newcommand{\thetako}{\pmb{\theta}^{(\iota+1)}}

\newcommand{\ds}{\displaystyle}

\newtheorem{mylem}{\textbf{Lemma}}

\newcommand{\la}{\langle}
\newcommand{\ra}{\rangle}

\newcommand{\Prf}{{\it Proof: \ }}

\newcommand{\clC}{{\cal C}}

\newcommand{\clR}{{\pmb{\cal R}}}

\newcommand{\wko}{\pmb{\text{w}}^{(\iota+1)}}

\newcommand{\wk}{\pmb{\text{w}}^{(\iota)}}

\newcommand{\bbC}{\mathbb{C}}

\allowdisplaybreaks
\begin{document}
	\title{Secure Communications for \emph{All Users} in Low-Resolution IRS-aided Systems Under Imperfect and Unknown CSI}
	\author{Monir Abughalwa 
		\IEEEmembership{Student Member, IEEE}, Diep N. Nguyen, \IEEEmembership{Senior Member, IEEE},\\ Dinh Thai Hoang,\IEEEmembership{Senior Member, IEEE},  Thang X. Vu, \IEEEmembership{Senior Member, IEEE}, \\ Eryk Dutkiewicz, \IEEEmembership{Senior Member, IEEE}, and Symeon Chatzinotas, \IEEEmembership{Fellow, IEEE}
		\thanks{Monir Abughalwa, Diep N. Nguyen, Dinh Thai Hoang, and Eryk Dutkiewicz are with School of Electrical and Data Engineering, University of Technology Sydney, Sydney, Australia (e-mail: monir.abughalwa@student.uts.edu.au; diep.nguyen@uts.edu.au; hoang.dinh@uts.edu.au; eryk.dutkiewicz@uts.edu.au).}
		\thanks{Thang X. Vu, and Symeon Chatzinotas are with the Interdisciplinary Centre for Security
			Reliability and Trust (SnT), Universite du Luxembourg, 
			Luxembourg, Luxembourg (e-mail:thang.vu@uni.lu; symeon.chatzinotas@uni.lu).}
	}

	\maketitle
	\begin{abstract}
		Provisioning secrecy for \emph{all users}, given the heterogeneity and uncertainty of their channel conditions, locations, and the unknown location of the attacker/eavesdropper, is challenging and not always feasible. This work takes the first step to guarantee secrecy for \emph{all users} where a low-resolution intelligent reflecting surface (IRS) is used to enhance legitimate users' reception and thwart the potential eavesdropper (Eve) from intercepting. In real-life scenarios, due to hardware limitations of the IRS's passive reflective elements (PREs), the use of a full-resolution (continuous) phase shift (CPS) is impractical. In this paper, we thus consider a more practical case where the phase shift (PS) is modeled by a low-resolution (quantized) phase shift (QPS) while addressing the phase shift error (PSE) induced by the imperfect channel state information (CSI). To that end, we aim to maximize the minimum secrecy rate (SR) among \emph{all users} by jointly optimizing the transmitter’s beamforming vector and the IRS’s passive reflective elements (PREs) under perfect/imperfect/unknown CSI. The resulting optimization problem is non-convex and even more complicated under imperfect/unknown CSI. To tackle it, we linearize the objective function and decompose the problem into sequential subproblems. When the perfect CSI is not available, we use the successive convex approximation (SCA) approach to transform imperfect CSI related semi-infinite constraints into finite linear matrix inequalities (LMI). We prove that our proposed algorithm converges to a locally optimal solution with low computational complexity thanks to our closed-form linearization approach. This makes the solution scalable for large IRS deployments.  Extensive simulations with practical settings show that our approach can ensure secure communication for \emph{all users} while the IRS's PREs are quantized and are affected by the PSE. 
	\end{abstract}
	
	\section{Introduction}
	\blfootnote{Preliminary results of this work are presented at the IEEE GLOBECOM Conference, 2024, \cite{10901599}.} Intelligent reflective surface (IRS) is a promising technology that has attracted paramount interest. IRSs comprise passive reflective elements (PRE) with a phase shift (PS) controller to tailor the reflected signals upon them \cite{nguyen2022leveraging}. By optimizing signal reflections, IRSs improve reception by creating favorably reflected multi-path signals at the receivers. Thanks to their cost-effective design and convenient deployment typically on the facades of high-rise buildings, IRSs hold immense potential for various applications, particularly in urban areas where line-of-sight channels between transmitters (Tx) and receivers (Rx) frequently face obstructions \cite{abughalwa2022finite}. Particularly, for the Internet of Things (IoT) devices that have limited computing capability and battery/energy, the IRS has gained paramount attention aiming to enhance both spectral and energy efficiency\cite{9896755}. 
	\vspace{-0.4cm}\subsection{Related Works and Motivations}
	Another potential application of IRSs is to enhance the security/privacy of users by purposely manipulating reflected signals from the Tx so as to facilitate the signal reception at legitimate users while maximizing the multi-user interference/degrading the signals at potential eavesdroppers. In \cite{zhou2021secure}, the authors studied the design of an IRS to maximize the legitimate user's secrecy rate (SR), which is defined as the difference between the rate of the legitimate channel and that of the channel from the transmitter to a potential eavesdropper. The authors studied SR maximization by jointly optimizing the beamforming vector at the transmitter and the IRS's PREs. A multi-input-single-output (MISO) IRS-aided system with the presence of multiple eavesdroppers was considered in \cite{hong2020robust}. The transmitter introduced artificial noise (AN) as a countermeasure to enhance user security.  Shi et al. in \cite{9913501} investigated the SR in a multi-input multi-output (MIMO) IRS-aided system with the presence of a single eavesdropper. The IRS was deployed to enhance the uplink transmission and the downlink energy transfer. The authors investigated the problem of SR maximization by jointly optimizing the transmit beamform vector, the downlink/uplink time allocation, and the energy transmit covariance matrix.

	When the channel state information (CSI) from the IRS to the users/receivers is unknown or imperfect, the authors in \cite{10256584} studied an IRS-aided multi-user system, where they formulated a secrecy sum rate (SSR) maximization problem with the eavesdropper's channel partially known to the receiver. In \cite{9402750}, the authors investigated an IRS-aided cognitive radio system when the eavesdropper's CSI is not available at Alice. The authors proposed a power minimization problem to guarantee users' secrecy by optimizing the beamforming vector and the IRS PREs, which were modeled by continuous PS. It is worth mentioning that most of the research in IRS's secrecy and signal-to-interference plus noise ratio (SINR) maximization assumes perfect phase estimation and/or full resolution/continuous phase shift (CPS) \cite{zhou2021secure,9180053}. However, assuming CPS of the IRS's PREs is not practical in real-life scenarios due to hardware limitations \cite{7510962}. In \cite{8683145}, the authors studied an IRS-aided system with a discrete/quantized phase shift (QPS). The authors proposed a transmission power minimization problem to achieve a certain user's SNR by jointly optimizing the beamforming vector and the discrete PREs. In \cite{8746155}, it was shown that using a 3-bit QPS can nearly achieve the same performance as a CPS. When the transmitter can obtain partial/imperfect CSI, the IRS's PREs are affected by phase shift error (PSE). The PSE was investigated in \cite{8869792}, where the IRS PREs PSE was presented in a large IRS system. 
	
	Motivated by the above, this paper aims to achieve secrecy for \emph{all users} under low-resolution IRS-aided systems with both perfect and imperfect CSI. To that end, we consider a popular use case where a low-resolution IRS is used to enhance/aid the signal reception at legitimate users (from a transmitter) in the presence of a potential eavesdropper. We then maximize the minimum SR by optimizing the transmit beamforming vector and the IRS's PREs. 
	
	When Alice can only obtain partial or erroneous knowledge of the IRS-to-Users/Eve's CSI, the problem becomes more challenging. Mathematically, imperfect CSI introduces non-convex constraints into the problem. To tackle it, we use slack variables to manage the coupling between the transmitter's beamforming vectors and the PREs PS within the objective function. We then apply the successive convex approximation (SCA) technique \cite{9197675}, and the $\mathcal{S}$-procedure \cite{boyd1994linear} to reformulate the semi-infinite constraints into linear matrix inequalities (LMI). \textcolor{black}{While the SCA framework combined with the $\mathcal{S}$-procedure transforms the semi-infinite constraints into an LMI, the resulting LMIs remain nonconvex. To address this, we employ a first-order Taylor approximation to linearize these non-convex components, enabling an efficient convex reformulation that can be solved at each iteration.} Finally, to handle the UMC of the IRSs, we employ the penalty convex-concave procedure (PCCP) \cite{9525400}. \textcolor{black}{Unlike other research that doesn't account for PSE \cite{9180053}, in this work, we leverage the PCCP algorithm to account for PSE in the IRS optimization.}
    For comparison purposes, we also consider the SSR maximization problem. When the eavesdropper's CSI is unknown to the transmitter, the IRS PREs modeled by  QPS suffer from PSE. To provide secrecy for \emph{all users}, we develop a power minimization problem while introducing a residual power that acts as an AN to lower the eavesdropper's SINR. Extensive simulations with practical settings show that by maximizing the minimum SR among all the users or minimizing transmit power and using AN, we can achieve a better chance of ensuring secure communications for \emph{all users} (subject to the location of the eavesdropper) even under imperfect CSI and low-resolution IRS. Where feasible, as expected, we observe that maximizing the minimum SR can achieve greater fairness among the users compared to the SSR maximization problem. 
	\vspace{-0.3cm}\subsection{Contributions}
	The main contributions are summarized as follows:
	\begin{itemize}
		\item  We study the SR guarantee for \emph{all users} and the impact of the low-resolution IRS in a multi-user downlink IRS-aided network. We formulate three optimization problems: maximizing the minimum SR, maximizing the SSR, and minimizing the transmission power, by jointly optimizing the transmitter's beamforming vector and the IRS's PREs. We then study the problems under both perfect and imperfect CSI cases.
		\item In the first case, we tackle the non-convex problem under perfect CSI. We linearize its objective function using mathematical approximations that yield a mathematically tractable approximate surrogate function. \textcolor{black}{To handle the UMC, and in contrast to conventional approaches such as semi-definite relaxation (SDR), we circumvent the computationally intractable UMC by directly optimizing the quantized PREs argument. Consequently, the PRE design reduces to an optimization over trigonometric functions, yielding a low-complexity, scalable solution for large-scale IRS deployments.} The proposed algorithm is analytically shown to converge to a locally optimal solution of the original non-convex problem.
		\item When only partial CSI is known to Alice, we leverage the SCA to tackle the imperfect CSI-related semi-infinite constraints to transform them into LMIs. \textcolor{black}{We then linearize the non-convex terms within these LMIs using a first-order Taylor approximation}. Next, we tackle the UMC using the PCCP algorithm, \textcolor{black}{explicitly accounting for PSE arising from channel uncertainty.}. We show that the proposed algorithm converges to, at least, a locally optimal solution of the non-convex problem. 
        \item In the final scenario, where the eavesdropper's CSI is unknown, we formulate a power minimization problem that ensures the users' data rates remain above a predefined quality-of-service (QoS) threshold. The residual transmit power is then allocated as AN to degrade the eavesdropper's SINR. The UMC is handled via PCCP while ensuring QoS constraints are met, \textcolor{black}{and accounts for the PSE}. The proposed algorithm is proven to converge to a locally optimal solution, effectively enhancing secrecy for all users.
		\item Eventually, we conduct extensive simulations to assess the impact of optimizing the IRS's PREs on users' SR. The results demonstrate that the proposed algorithms ensure secure communication for \emph{all users}. In contrast, approaches that focus on maximizing the SSR, i.e. \cite{9110587}, fail to guarantee secrecy for \emph{all users}.
	\end{itemize}
	The remainder of this paper is organized as follows. The system model is discussed in Sections \ref{sec2}. Then, Sections \ref{sec3}, \ref{sec4}, and \ref{secAN2} present the problem and corresponding solutions under prefect/imperfect/unknown CSI, respectively. The extensive simulations and discussion are in Section \ref{sec5}. Finally, Section \ref{conc} concludes the paper and discusses potential future directions.
	
	This paper uses the following notation: bold letters denote vectors and matrices. $\pmb{I}_M$ denotes an M dimensional identity matrix. Diag($n_1,\dots,n_n$) denotes the diagonal matrix with diagonal entries of $\{n_1,\dots,n_n\}$. The symbols $\Re$ and $\mathbb{C}$ represent the real and the complex field, respectively. $\clC(0,\bar{z})$ denotes the circular Gaussian random variable with zero mean and variance $\bar{z}$. For matrices $\mathbf{C}$ and $\mathbf{D}$, $\la \mathbf{C},\mathbf{D} \ra \triangleq \mbox{trace}\left(\mathbf{C}^H \mathbf{D}\right)$. For matrix $\mathbf{A}$, $\Re\left\{\mathbf{A}\right\}$ denotes the real part, $\la \mathbf{A}\ra \triangleq\mbox{trace}(\mathbf{A})$, the symbol $\|\mathbf{A}\|_1$ denotes the 1-norm, $\|\mathbf{A}\|$ denotes the Frobenius norm, $\mathbf{A}^*$ denotes the conjugate, $\mathbf{A}^H$ denotes the Hermitian (conjugate transpose), $\lambda_{\text{max}}\left(\mathbf{A}\right)$ denotes the maximal eigenvalue, $\angle(\mathbf{A})$ denotes its argument, and $\mathbf{A} \succeq 0$ means positive semi-definite.  
	\section{{System Model}} \label{sec2}
	\begin{figure}[!htb]
		\centering
		\includegraphics[width=0.33\textwidth]{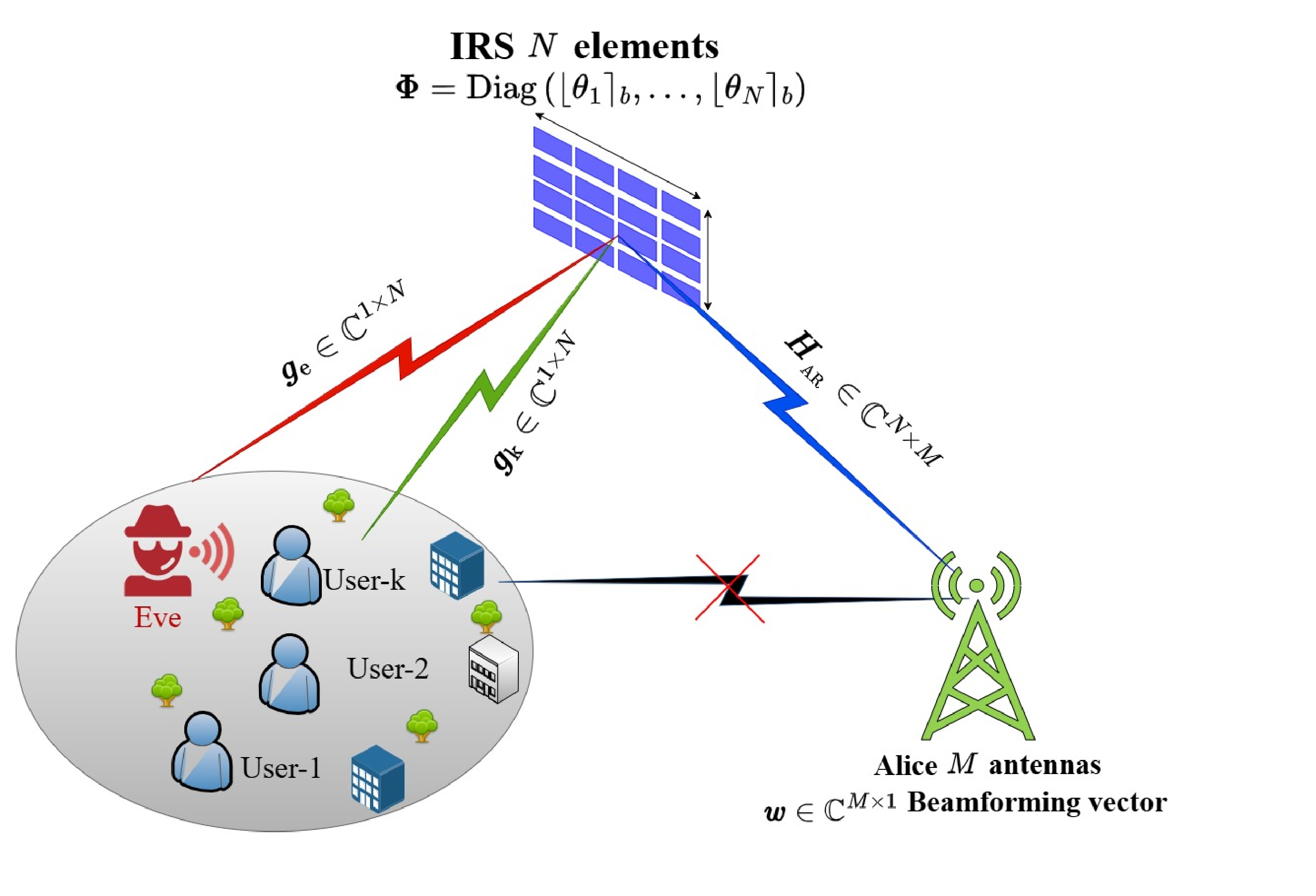}
		\caption{System Model.}
		\label{s1}
	\end{figure}
	We consider a downlink system aided by a low-resolution IRS, as illustrated in Fig. \ref{s1}. An $M$-antenna transmitter (Alice) transmits confidential messages to $\mathcal{K}$ single antenna legitimate users, while a single antenna eavesdropper (Eve) attempts to wiretap the transmission. To support the transmission between Alice and the users,
	an IRS with $N$ PREs is deployed (e.g., at the facade of a building) to establish a link between Alice and the users. Let $k \triangleq \{1,2, \dots, K\}$ denote the set of legitimate users and $e$ denote Eve in the system. The direct channel from Alice to the users/Eve is denoted by
	$\pmb{h}_{A_i} \triangleq \sqrt{\beta_{A_i}} \tilde{\pmb{h}}_{A_i} $, where $\tilde{\pmb{h}}_{A_i}$ is modelled by Rician fading, and $\sqrt{\beta_{A_i}}$ is the large scale fading from the Alice-to-users/Eve link, $\pmb{H}_\text{AR}\triangleq \sqrt{\beta_{_{\text{{AR}}}}} \tilde{\pmb{H}}_{_\text{{AR}}}\in \mathbb{C}^{N \times M}$ denote the channel from Alice to the IRS, while $\tilde{\pmb{H}}_{\text{{AR}}}$ is modeled by Rician fading, and $\sqrt{\beta_{_{\text{{AR}}}}}$ is the large scale fading factor of the Alice-to-IRS link. The channel from the IRS to user $k$ and to Eve is modeled by ${\pmb{g}}_i\triangleq  \sqrt{\beta_{\text{R}i}} \tilde{{\pmb{g}}}_i \clR^{1/2}_{\text{R}i}\in \mathbb{C}^{1\times N}$, where $i \in \{k,e\}$, $\beta_{\text{R}i}$ is the large-scale fading factor of the IRS-to-$i$ \cite{kammoun2020asymptotic}, $\tilde{{\pmb{g}}}_i$ is modeled by Rician fading, and $\clR_{\text{R}i} \in \bbC^{N\times N}$ is the IRS elements' spatial correlation matrix \cite{kammoun2020asymptotic}. For the confidential message intended for user $k$, $s_k$, the signal received by user $k$ and Eve {with corresponding to the intended user}, respectively, can be expressed as: 
	\begin{align} \label{chkE}
		y_i \triangleq \pmb{h}_{i}(\btheta) {\sum}_{k=1}^{\mathcal{K}} \pmw_k s_k +n_i, ~i\in \{k,e\},
	\end{align}
	where  $\pmb{h}_{i}(\btheta) \in \mathbb{C}^{1 \times M} $ is the cascaded channel gain from Alice to $i \in \{k,e\}$, $\pmw_k \in \mathbb{C}^{M \times 1}$ represents the beamforming vector applied for user $k$,  $\btheta\triangleq(\theta_1,\dots, \theta_N)^T\in [0,2\pi)^N$ denotes the IRS's PREs PS vector, $n_i$ is the zero-mean Additive White Gaussian Noise (AWGN) with power $\sigma_i$ for $i \in \{k,e\}$. The cascade channel gain $\pmb{h}_{i}(\btheta)$ from Alice to $i \in \{k,e\}$, can be expressed in terms of the direct channel gain from Alice to the IRS, and the channel gain from the IRS to the user or Eve as follows: 
	\vspace{-0.1cm}\begin{align} \label{h1} 
		\pmb{h}_{i}(\btheta)\triangleq {{\pmb{g}}}_i \pmb{\Phi} {\pmb{H}_\text{AR}} \triangleq {{\pmb{g}}}_i {\sum}_{n=1}^{N} \exp(j \theta_n) \Xi_n {\pmb{H}_\text{AR}},
	\end{align}
	where  $\pmb{\Phi} \triangleq \text{Diag}(e^{j \btheta})$, $\pmb{\Xi}_n$ is an $N \times N$ matrix with all zeros except for its $(n,n)$ entry which is 1.
	\begin{figure*}[!t] 
		\begin{align} 
			C_e(\pmw,\lfloor\thetak\rceil_{b}) &\geq q_{1,k_{e}}^{(\iota)}+2\Re \sum\limits_{j=1, j \neq k}^{K}\left\{\left\la \pmb{m}_{j,k_{e}}^{(\iota)},\pmw_j\right\ra\right\} -\frac{\rho_{e}^{(\iota)}}{1+\rho_{e}^{(\iota)}}\left(\sum\limits_{j=1,j \neq k}^{K}|\pmb{h}_{e}(\lfloor\thetak\rceil_{b})\pmw_j|^2\right) - \frac{1}{\upsilon_{e}^{(\iota)}}\sum\limits_{j=1}^{K}|\pmb{h}_{e}(\lfloor\thetak\rceil_{b})\pmw_j|^2, \label{CEtot1} \tag{15} \\
			\pmb{\psi}^{(\iota)}_k&\triangleq \sum\limits_{j=1}^{\mathcal{K}}n_{1,k}^{(\iota+1)}\pmb{h}^H_{k}(\lfloor\thetak\rceil_{b})\pmb{h}_{k}(\lfloor\thetak\rceil_{b})+\frac{\rho_e^{(\iota)}}{1+\rho_e^{(\iota)}}\sum\limits_{j=1,j \neq k}^{\mathcal{K}}\pmb{h}^H_{e}(\lfloor\thetak\rceil_{b})\pmb{h}_{e}(\lfloor\thetak\rceil_{b}) +  \frac{1}{\upsilon_{e}^{(\iota)}}\sum_{j=1}^{\mathcal{K}}\pmb{h}^H_{e}(\lfloor\thetak\rceil_{b})\pmb{h}_{e}(\lfloor\thetak\rceil_{b}),  \label{qq2} \tag{17} \\
			{C_k}(\wko,\lfloor\btheta\rceil_b) & \geq {q}_{1,k}^{(\iota+1)}+2\Re\left\{{\sum}_{n=1}^{N}{\pmb{m}}^{(\iota+1)}_{k,k}(n) e^{j\lfloor\theta_n\rceil_b} \right\}+ \left(e^{j \lfloor\btheta\rceil_b}\right)^H {\pmb{\varphi}}_{1,k}^{(\iota+1)} e^{j \lfloor\btheta\rceil_b}, \label{LBBob2x} \tag{19}\\
			{q}_{1,k}^{(\iota+1)} &\triangleq C_k(\wko,\lfloor\thetak\rceil_b)-\gamma_k(\wko,\lfloor\thetak\rceil_b)-\sigk {n}_{1,k}^{(\iota+1)}, \label{qck} \tag{20}\\
			{n}_{1,k}^{(\iota+1)} &\triangleq{1}/{\rho_{k}^{(\iota+1)}}-{1}/{\upsilon_{_{k}}^{(\iota+1)}}. \label{nck} \tag{21}
		\end{align} 
		\hrulefill 
	\end{figure*}
    
	Since the IRS is typically deployed at a known fixed location, such as the facade of a high-rise building, the CSI between Alice and the IRS can be accurately estimated by exploiting the angles of arrival and departure \cite{9110587}, or by employing advanced channel estimation techniques \cite{9501057,9373363}. However, obtaining the IRS-users' reflected channel’s CSI is considerably more challenging due to the passive nature of the IRS and the mobility and dynamic environment of the users  \cite{8579566}. For Eve, pinpointing its location or its accurate CSI is hardly possible. To address these uncertainties, we adopt a bounded CSI error model wherein the reflected channels from the IRS to the users and Eve are expressed as \cite{9266086}:	 
	\begin{subequations} \label{ICSIcha}
		\begin{alignat}{3} 
			{\pmb{g}}_{i}&\triangleq \widehat{{\pmb{g}}}_{i} + \Delta{\pmb{g}}_{i}, i \in \{k,e\}, \label{ch2-h1}	 \\            
			\omega_i &\triangleq \{\|\Delta {\pmb{g}}_{i}\|_2 \leq \xi_{i}\}, i \in \{k,e\}, \label{ch2-h2}
		\end{alignat}
	\end{subequations}
	where $\hat{{\pmb{g}}}_{i}$ denotes the (imperfect) estimated channel vector, $\Delta{\pmb{g}}_{i}$ represents the channel estimation error of the corresponding estimation, $\omega_i$ is a set for all possible channel estimation errors, and $\xi_{i}$ is the radii of the uncertainty regions as known to Alice. Thus, \eqref{h1} can be reformulated as:
	\begin{align} \label{chaCSIdef}
		\widehat{\pmb{h}}_{i}(\btheta)\triangleq \left(\widehat{{{\pmb{g}}}}_i+ \Delta{\pmb{g}}_{i}\right) \pmb{\Phi} {\pmb{H}_\text{AR}}.
	\end{align}
	Hence, the signal-to-interference-plus-noise ratio (SINR) of the received signal at the user-$k$ and Eve, under perfect/imperfect CSI, can be expressed as:
	\begin{align}
		\gamma_i(\pmw,\btheta)&\triangleq{|\pmb{h}_{i}(\btheta)\pmw_k|^2}/{\rho_i(\pmw,\btheta)}, ~ i \in \{k,e\}, \label{SINRPCSI}\\
		\widehat{\gamma}_i(\pmw,\btheta)&\triangleq {|\widehat{\pmb{h}}_{i}(\btheta)\pmw_k|^2}/{\widehat{\rho}_i(\pmw,\btheta)}, ~ i \in \{k,e\}, \label{SINRICSI} 
	\end{align}
	where  $\rho_i(\pmw,\btheta)\triangleq {\sum}_{j=1,j \neq k}^{\mathcal{K}}|\pmb{h}_{i}(\btheta)\pmw_j|^2+\sigma_i$, and  $\widehat{\rho}_i(\pmw,\btheta)\triangleq {\sum}_{j=1,j \neq k}^{\mathcal{K}}|\widehat{\pmb{h}}_{i}(\btheta)\pmw_j|^2+\sigma_i$.
	However, in a low-resolution IRS-aided system with imperfect CSI from the IRS to the users/Eve, the IRS PREs are affected by PSE \cite{8869792}. Thus, in the following, we consider three cases to model the IRS's PREs as:
	\begin{itemize}
		\item The CPS adjustment at the IRS; this is the ideal case when the IRS PREs can be modeled by a continuous PS.
		\item  The low-resolution/QPS at the IRS; where the PS can be modeled by a uniform distribution in $\lfloor \btheta \rceil_b \in [\frac{-\pi}{2^b}, \frac{\pi}{2^b})$, where $b$ is the number of bits in the QPS modulation.
		\item with imperfect/unknown CSI, and when only a discrete set of $2^b$ phases can be configured; this leads to a PSE ${\epsilon}_R$ which can be approximated by a uniform distribution in $ [-2^{-b} \pi,2^{-b} \pi)$, hence, \eqref{chkE} can be reformulated as:
		\begin{align} \label{checkerror} 
			\pmb{h}_{i}(\lfloor\btheta\rceil_{b})\triangleq  {{\pmb{g}}}_i \text{Diag}(e^{j (\lfloor\btheta\rceil_b+\epsilon_R)}) {\pmb{H}_\text{AR}}.
		\end{align}
	\end{itemize}
	
	In the sequel, we consider both perfect and imperfect CSI scenarios for the IRS to the users and Eve. In the latter, only partial or estimated CSI $\widehat{{\pmb{g}}}_{i}$ is available. To capture different levels of CSI imperfection, one can vary the magnitude of $\Delta{\pmb{g}}_{i}$, i.e., the radii $\xi_{i}$ of the uncertainty region. In a more extreme scenario, we assume that Alice has access only to imperfect IRS-to-users CSI, while the CSI of Eve remains completely unknown to Alice.
	
	Under the presence of Eve, the closed-form expression for the SR of the user-$k$ under perfect/imperfect/unknown IRS to users/Eve CSI is defined as \cite{bloch2008wireless}:
	\begin{align} 
		\mbox{SR}_{k}(\pmw,\lfloor\btheta\rceil_{b})& \triangleq [C_k(\pmw,\btheta)-C_e(\pmw,\btheta)]^+, \label{Cstot1}\\
		\widehat{\mbox{SR}}_{k}(\pmw,\lfloor\btheta\rceil_{b})& \triangleq [\widehat{C}_k(\pmw,\btheta)-\widehat{C}_e(\pmw,\btheta)]^+,  \label{Cstot2}
	\end{align}
	Where $[x]^+ \triangleq \text{max}[0,x]$, the user-$k$ data rate, and the eavesdropping rate under perfect/imperfect CSI are defined as \cite{bloch2008wireless}: 
	\vspace{-0.3cm}\begin{align} 
	C_i(\pmw,\lfloor\btheta\rceil_{b})&\triangleq\ln\left(1+\gamma_i(\pmw,\btheta)\right), i \in \{k,e\}, \label{ckedef1}  \\
		\widehat{C}_i(\pmw,\lfloor\btheta\rceil_{b})&\triangleq\ln\left(1+\widehat{\gamma}_i(\pmw,\btheta)\right), i \in \{k,e\}. \label{ckedef2} 
	\end{align}

	\section{Minimum user’s SR maximization under perfect CSI} \label{sec3}
	\subsection{MAXMIN SR under perfect CSI}\label{sec3-p1}
	In this section, we first address the problem of maximizing the minimum SR among all the users, under the perfect CSI assumption, to guarantee secure communications for \emph{all users}. The optimization problem can be formally stated as 
	\begin{subequations} \label{P1}
		\begin{alignat}{3} 
			&(\mathcal{P}1):&& ~ \underset{\pmw,\lfloor\btheta\rceil_b}{\max}~\underset{k \in \mathcal{K}}{\min} ~ \mbox{SR}_{k}(\pmw,\lfloor\btheta\rceil_b), \label{P1-1} \\            
			&~\text{s.t.} && ~ {\sum}_{k = 1}^{\mathcal{K}}\Vert \pmw_k \Vert^2 \leq P_T,\label{P1-2} \\
			& && ~ |e^{(j \lfloor\btheta\rceil_b)}| = 1,  \label{P1-3}
		\end{alignat}
	\end{subequations}
	where $P_T$ is the transmit power. \eqref{P1-2} captures the sum of the transmitted power constraint and \eqref{P1-3} captures the UMC of the PREs' PS.
	
	The optimization problem ($\mathcal{P}1$) is nonconvex since the objective function \eqref{P1-1} is {nonlinear and} not concave, and the UMC (\ref{P1-3}) is nonconvex. {Moreover, the coupling between $\pmw$ and $\lfloor\btheta\rceil_b$ within the objective function $\eqref{P1-1}$ further complicates solving problem ($\mathcal{P}1$). To address this coupling, one can employ the AO technique \cite{nguyen2022leveraging}. Specifically, at iteration $(\iota)$, the feasible point $(\wk,\lfloor\thetak\rceil_b)$ is generated from ($\mathcal{P}1$) by solving two sub-problems. Firs, we tackle the subproblem by optimizing $\pmw$ with a fixed $\lfloor\btheta\rceil_b$,}
	\vspace{-0.3cm}\begin{equation*} 
		(\mathcal{P}1.1): ~ \underset{\pmw}{\max} ~ \underset{k}{\min} ~ \mbox{SR}_k(\pmw,\lfloor\thetak\rceil_b), ~ \text{s.t.} ~ (\ref{P1-2}),
	\end{equation*} 
	then, for a fixed $\pmw$, we optimize $\lfloor\btheta\rceil_b$ by solving the following subproblem,
	\begin{equation*} 
		(\mathcal{P}1.2): ~ \underset{\lfloor\btheta\rceil_b}{\max} ~ \underset{k}{\min} ~ \mbox{SR}_k(\wko,\lfloor\btheta\rceil_b), ~ \text{s.t.} ~ (\ref{P1-3}).
	\end{equation*}
	However, the AO approach requires solving two subproblems $(\mathcal{P}1.1)$ and $(\mathcal{P}1.2)$, which are computationally demanding, particularly due to the large number of PREs of the IRS. To address this, we propose a linearization method employing mathematically tractable approximation functions, leading to a computationally efficient algorithm that is able to handle a large number of IRSs' PREs.
	\subsubsection{Sub-Problem for Optimizing the Beamforming Vectors} \label{subsec1}
	We fix the quantized $\lfloor\btheta\rceil_b$ given $\wk$ and solve the problem ($\mathcal{P}1.1$) to obtain $\wko$ satisfying $\mbox{SR}_k(\wko,\lfloor\thetak\rceil_b) > \mbox{SR}_k(\wk,\lfloor\thetak\rceil_b)$. 
	We begin by linearizing the objective function in \eqref{P1-1}, which comprises two parts: the data rate of user-$k$ and Eve's negative eavesdropping rate. First, we transform the user-$k$'s data rate into a linear form. Specifically, applying the inequality \eqref{fund1} in Appendix \ref{AppA}, we define $\pmb{\Lambda}\triangleq\pmb{h}_{k}(\lfloor\btheta\rceil_b^{(\iota)})\pmw_k$, $\pmb{\digamma}\triangleq\rho_k(\pmw,\lfloor\thetak\rceil_b)$, $\hat{\Lambda}\triangleq\pmb{h}_{k}(\lfloor\btheta\rceil_b^{(\iota)})\pmw_k^{(\iota)}$ and
	$\hat{\digamma} \triangleq \rho_k(\pmw^{(\iota)},\lfloor\btheta\rceil_b^{(\iota)})$, hence, the users' data rate \eqref{ckedef1} can be written as: \setcounter{equation}{12} 
	\begin{equation} \label{LBBob} 
		C_k(\pmw,\lfloor\thetak\rceil_b)\geq q_{1,k}^{(\iota)}+2\Re\left\{\left\la \pmb{m}_{k,k}^{(\iota)},\pmw_k\right\ra\right\}-n_{1,k}^{(\iota)}\upsilon_{_{k}}^{(\iota)},
	\end{equation}
	where, 
	\vspace{-0.3cm}\begin{align*}
		\upsilon_{k}^{(\iota)}&\triangleq {\sum}_{j=1}^{\mathcal{K}}|\pmb{h}_{k}(\lfloor\thetak\rceil_b)\pmw_j|^2+\sigk, \\
		q_{1,k}^{(\iota)} &\triangleq C_k(\pmw,\lfloor\thetak\rceil_b)-\gamma_k(\pmw,\lfloor\thetak\rceil_b)-\sigk n_{1,k}, \\
		\pmb{m}^{(\iota)}_{k,k}&\triangleq {(\pmb{h}_{k}^H (\lfloor\thetak\rceil_b)\pmb{h}_{k}(\lfloor\thetak\rceil_b)\pmw_k)}/{(\rho_{_{k}}^{(\iota)})}, \\
		n_{1,k}^{(\iota)}&\triangleq{1}/\rho_{_{k}}^{(\iota)}-{1}/\upsilon_{_{k}}^{(\iota)}.
	\end{align*}
	
	Second, we tackle the nonlinear Eve's negative eavesdropping rate, which can be expressed as:
	\vspace{-0.3cm}\begin{align}\label{EveSRt1}
		-\ln(1+\gamma_e(\pmw,\lfloor\thetak\rceil_b))&\triangleq \overset{a_1}{\overbrace{\ln(1+\rho_{e}^{(\iota)})}}- \overset{a_2}{\overbrace{\ln(1+\upsilon_e^{(\iota)}})}, 
	\end{align}
	{where $\upsilon_{e}^{(\iota)}\triangleq {\sum}_{j=1}^{\mathcal{K}}|\pmb{h}_{e}(\lfloor\thetak\rceil_b)\pmw_j|^2+\sige$. 
	To linearize the term ($a_1$) in \eqref{EveSRt1}, we adopt the idea in \cite{niu2022joint}. The term ($a_1$) in \eqref{EveSRt1} can be linearized by introducing the auxiliary function $\pmb{z}\triangleq\rho_e^{(\iota)}$, which is then incorporated into inequality \eqref{fund2} in Appendix \ref{AppA}. The term ($a_2$) in \eqref{EveSRt1} is linearized by defining $\pmb{\Upsilon}\triangleq \upsilon_e^{(\iota)}$ and substituting it in the inequality \eqref{fund3} in Appendix \ref{AppA}.}
	\begin{figure*}[!t] 
		\begin{align} 
			C_e(\wko,\lfloor\btheta\rceil_b)&\geq q_{1,k_{e}}^{(\iota+1)}+2\mathfrak{R}\left\{{\sum}_{n=1}^{N} {\pmb{m}}^{(\iota+1)}_{k_{e}}(n) e^{(j \lfloor\theta_n\rceil_b)} \right\}+\left(e^{j \btheta}\right)^H {\pmb{\varphi}} _{1,k_{e}}^{(\iota+1)} e^{j \lfloor\btheta\rceil_b} + \left(e^{j \lfloor\btheta\rceil_b}\right)^H {\pmb{\varphi}} _{k_{e},j}^{(\iota+1)} e^{j \lfloor\btheta\rceil_b}, \label{LBEve2x} \tag{22} \\
			{q}_k^{(\iota+1)}&\triangleq {q}_{1,k}^{(\iota+1)}+q_{1,ke}^{(\iota+1)}- \left(e^{j \lfloor\thetak\rceil_b}\right)^H\tilde{{\pmb{\varphi}}}_{k}^{(\iota+1)}e^{j\lfloor\thetak\rceil_b}-2\lambda_{\text{max}}\left(\tilde{{\pmb{\varphi}}}_{k}^{(\iota+1)}\right) N, \label{qkth} \tag{24}\\
			{\pmb{m}}^{(\iota+1)}_k(n)&\triangleq\hat{\pmb{m}}^{(\iota+1)}_{k,k}(n)+\tilde{\pmb{m}}^{(\iota+1)}_{k}(n)+ {\sum}_{m=1}^{N}e^{-j \lfloor\theta_m^{(\iota)}\rceil_b} \tilde{{\pmb{\varphi}}}_{k}^{(\iota+1)}(m,n)+\lambda_{\text{max}}\left(\tilde{{\pmb{\varphi}}}_{k}^{(\iota+1)}\right), \label{mkth} \tag{25}\\
			\lfloor\thetako\rceil_b &\triangleq \arg \underset{\lfloor\thetako\rceil_b \in \left\{\lfloor\btheta\rceil_b^{(\iota+1),k}, k=1,\dots,K\right\}}{\max} \mbox{SR}_k(\wko,\lfloor\btheta\rceil_b). \label{solth} \tag{28}
		\end{align} 
		\hrulefill 
	\end{figure*}
	
	Hence, Eve's eavesdropping rate can be expressed as in \eqref{CEtot1}, where, 
	\vspace{-0.3cm}\begin{align*}
		q_{1,k_{e}}^{(\iota)} &\triangleq \ln\left(\rho_{e}^{(\iota)}\right)-\rho_{e}^{(\iota)}-\ln\left(\upsilon_{e}^{(\iota)}\right)+n_{1,k_{e}}^{(\iota)}+1, \\
		n_{1,k_{e}}^{(\iota)} &\triangleq\left(-{\rho_{e}^{(\iota)}}/{(1+\rho_{e}^{(\iota)})}-{1}/{\upsilon_{e}^{(\iota)}}\right)\sige, \\
		\pmb{m}_{j,k_{e}}^{(\iota)} &\triangleq \pmb{h}_{e}^H(\thetak)\pmb{h}_{e}(\thetak) \pmw_j^{(\iota)}.
	\end{align*}
	
	By substituting (\ref{LBBob}) and (\ref{CEtot1})  into (\ref{Cstot1}), we can express the {approximate surrogate} SR as: \setcounter{equation}{15}
	\begin{align} \label{SRLBR}
		\tilde{\mbox{SR}}_{k}(\pmw,\lfloor\thetak\rceil_b) &\geq q_{k}^{(\iota)}+2\Re\left\{\left\la \pmb{m}_{k}^{(\iota)},\pmw_k\right\ra\right\}-(\pmw_k)^H \pmb{\psi}^{(\iota)}_k \pmw_k,
	\end{align}
	where $\pmb{\psi}^{(\iota)}_k$ is defined in \eqref{qq2}, and 
	\begin{align*}
		q_{k}^{(\iota)}&\triangleq q_{1,k}^{(\iota)}+q_{1,ke}^{(\iota)}+n_{1,ke}^{(\iota)}, \\
		\pmb{m}_{k}^{(\iota)}&\triangleq \pmb{m}_{k,k}^{(\iota)}+{\sum}_{j=1,j \neq k}^{\mathcal{K}}\tilde{\pmb{m}}_{j,k}^{(\iota)},\\
		\tilde{\pmb{m}}_{j,k}^{(\iota)}&\triangleq \pmb{h}_{e}^H(\lfloor\thetak\rceil_b)\pmb{h}_{e}(\lfloor\thetak\rceil_b) \pmw_j. 
	\end{align*}
	
	Finally, with (\ref{SRLBR}), problem ($\mathcal{P}1.1$) can be recast as \setcounter{equation}{17}
	\begin{subequations} \label{solw}
		\begin{alignat}{3} 
			&(\mathcal{P}1.3): && \max_{\Gamma, \pmw} ~{\Gamma}, \label{solw-1} \\            
			&\text{s.t.} && \Gamma \leq \tilde{\mbox{SR}}_{k}(\pmw,\lfloor\thetak\rceil_b), ~\forall k, \label{solw-12} \\
			& && \eqref{P1-2}, \label{solw-12x}
		\end{alignat}
	\end{subequations}
	where $\Gamma$ is an auxiliary variable as the lower bound of the SR. Problem ($\mathcal{P}1.3$) is an SDP problem characterized by the linearized objective function \eqref{SRLBR}. This optimization problem can be efficiently solved using standard convex optimization solvers, such as the interior-point method or the CVX toolbox \cite{grant2014cvx}.
	%
	%
	%
	\subsubsection{Sub-Problem for Optimizing the PREs} \label{subsec2}
	Similarly, by fixing $\pmw$, we seek $\lfloor\btheta\rceil_b^{(\iota+1)}$ such that, $\mbox{SR}_k(\wko_k,\lfloor\thetako\rceil_{b}) > \mbox{SR}_k(\wko_k,\lfloor\thetak\rceil_b).$ Following the same method in the previous section, and using the inequality \eqref{fund1} in Appendix \ref{AppA}, the user's data rate lower bound approximation can be obtained. \begin{algorithm} 
		\caption{Proposed Iterative Algorithm for Solving Problem $(\mathcal{P}1)$} \label{alg1}
		\begin{algorithmic}[1]
			\State \textbf{Initialize:} $(\pmb{\text{w}}^{(1)},\lfloor\btheta\rceil_b^{(1)})$, convergence tolerance $\epsilon_t > 0$,  and Set $\iota=1$. 
			\State \textbf{Repeat} 
			\State Update $\wko$ by (\ref{solw}), and $\thetako$ by (\ref{solth});
			\State \textbf{if} $\frac{|\underset{k}{\min} \mbox{SR}_{k}(\wko,\lfloor\thetako\rceil_{b})-\underset{k}{\min} \mbox{SR}_{k}(\wk,\lfloor\thetak\rceil_{b})|}{\underset{k}{\min} \mbox{SR}_{k}(\wk,\lfloor\thetak\rceil_{b})} \leq \epsilon_t$.
			\State \textbf{Then} $\lfloor\thetak\rceil_{b}\leftarrow \thetako$, $\wk\leftarrow \wko$ and terminate. 
			\State \textbf{Otherwise} $\iota\leftarrow \iota+1$ and continue.
			\State \textbf{Output} $(\wk_k, \thetak)$.
		\end{algorithmic}
	\end{algorithm}
    Accordingly, the user's data rate can be expressed as in \eqref{LBBob2x}, where ${q}_{1,k}^{(\iota+1)}$ and ${n}_{1,k}^{(\iota+1)}$ are given in \eqref{qck} and \eqref{nck}, respectively, and,
    		
	\begin{alignat*}{2}
		&\pmb{m}^{(\iota+1)}_{k,k}(n) &&\triangleq \hat{\pmb{m}}^{(\iota+1)}_{k,k}(n) / \rho_{k}^{(\iota+1)}, \\
		&\hat{\pmb{m}}^{(\iota+1)}_{k,k}(n) &&\triangleq \left(\wko_k \right)^H \pmb{h}_{_{k}}^H(\lfloor\thetak\rceil_b) {\pmb{g}}_k \pmb{\Xi}_n \pmb{H}_{_{\text{AR}}} \wko_k, \\
		&\pmb{\varphi}_{1,k}^{(\iota+1)} &&\triangleq -{n}_{1,k}^{(\iota+1)} {\sum}_{j=1}^{\mathcal{K}} \pmb{\varphi}_{k,j}^{(\iota+1)}, \\
		&\pmb{\varphi}_{k,j}^{(\iota+1)} &&\triangleq \left(\pmb{\aleph}_{k,j}^{(\iota+1)}(n)\right)^* \pmb{\aleph}_{k,j}^{(\iota+1)}(m),~ n \in N, m \in N, \\
		&\pmb{\aleph}_{k,j}^{(\iota+1)}(n) &&\triangleq {\pmb{g}}_k \pmb{\Xi}_n \pmb{H}_{_{\text{AR}}} \wko_j.
	\end{alignat*}
	Similarly, we can express Eve's eavesdropping rate as in \eqref{LBEve2x}, where, 
	\begin{alignat*}{2}
		&q_{1,k_{e}}^{(\iota+1)} &&\triangleq \ln\left(\rho_{e}^{(\iota+1)}\right) - \rho_{e}^{(\iota+1)} - \ln\left(\upsilon_{e}^{(\iota+1)}\right)+n_{1,k_{e}}^{(\iota+1)} + 1, \\
		&n_{1,k_{e}}^{(\iota+1)} &&\triangleq \left(-{\rho_{e}^{(\iota+1)}}/{1+\rho_{e}^{(\iota+1)}} - {1}/{\upsilon_{e}^{(\iota+1)}}\right) \sige, \\
		&\pmb{m}^{(\iota+1)}_{k_e}(n) &&\triangleq {\sum}_{j=1,j \neq k}^{\mathcal{K}} \pmb{m}^{(\iota+1)}_{j,k_{e}}(n), \\
		&\pmb{m}^{(\iota+1)}_{j,k_{e}}(n) &&\triangleq \left(\wko_k \right)^H \pmb{h}_{e}^H(\lfloor\thetak\rceil_b) {\pmb{g}}_{e} \pmb{\Xi}_n \pmb{H}_{_{\text{AR}}} \wko_k, \\
		&\pmb{\varphi}_{1,k_{e}}^{(\iota+1)} &&\triangleq \left(\frac{-\rho_{e}^{(\iota+1)}}{(1+\rho_{e}^{(\iota+1)})} - \frac{1}{(\sige+\upsilon_{e}^{(\iota+1)})}\right) \left(\sum\limits_{j=1}^{\mathcal{K}} \pmb{\varphi}_{k_{e},j}^{(\iota+1)} \right), \\
		&\pmb{\varphi}_{k_{e},j}^{(\iota+1)} &&\triangleq \left(\pmb{\aleph}_{k_{e},j}^{(\iota+1)}(n)\right)^* \pmb{\aleph}_{k_{e},j}^{(\iota+1)}(m), \\
		&\pmb{\aleph}_{k_{e},j}^{(\iota+1)}(n) &&\triangleq {\pmb{g}}_e \pmb{\Xi}_n \pmb{H}_{_{\text{AR}}} \wko_j.
	\end{alignat*}
	By combining (\ref{LBBob2x}) and (\ref{LBEve2x}) into (\ref{Cstot1}), we can express the closed-form SR as: \setcounter{equation}{22}
    \begin{algorithm}
		\caption{Proposed AO Algorithm for Solving Problem $(\mathcal{P}2)$} \label{alg2}
		\begin{algorithmic}[1]
			\State \textbf{Initialize:} $(\pmb{\text{w}}^{(1)},\lfloor\btheta^{(1)}\rceil_{b})$, convergence tolerance $\epsilon_t > 0$,  and Set $\iota=1$. 
			\State \textbf{Repeat} 
			\State Update $\wko$ by (\ref{sr3}), and $\lfloor\btheta\rceil_b$ by (\ref{solth2}); 
			\State \textbf{if} $\frac{|{\sum}_{k=1}^{\mathcal{K}} \mbox{SR}_{k}(\wko,\lfloor\thetako\rceil_b)-{\sum}_{k=1}^{\mathcal{K}} \mbox{SR}_{k}(\wk,\thetak)|}{{\sum}_{k=1}^{\mathcal{K}}\mbox{SR}_{k}(\wk,\lfloor\thetak\rceil_{b})} \leq \epsilon_t$
			\State \textbf{Then} $\lfloor\thetak\rceil_{b}\leftarrow \lfloor\btheta\rceil_b$, $\wk\leftarrow \wko$ and terminate. 
			\State \textbf{Otherwise} $\iota\leftarrow \iota+1$ and continue.
			\State \textbf{Output} $(\wk_k, \lfloor\thetak\rceil_{b})$.
		\end{algorithmic}
	\end{algorithm} 
	\begin{align} \label{SRtheta}
		\mbox{SR}_k(\wko,\lfloor\btheta\rceil_b) \geq {q}_k^{(\iota+1)}+ 2{\sum}_{n=1}^{N}\Re\left\{{\pmb{m}}^{(\iota+1)}_k(n) e^{j \lfloor\theta_n\rceil_b}\right\},
	\end{align}
	where ${q}_k^{(\iota+1)}$ is expressed in \eqref{qkth}, ${\pmb{m}}^{(\iota+1)}_k(n)$ is expressed in \eqref{mkth}, and $\pmb{\varphi}_{k}^{(\iota+1)} \triangleq \pmb{\varphi}_{1,k}^{(\iota+1)}+\pmb{\varphi}_{1,k_{e}}^{(\iota+1)}+\pmb{\varphi}_{k_{e},j}^{(\iota+1)}.$
	
    With \eqref{SRtheta}, we formulate the following optimization problem to obtain $\lfloor\thetako\rceil_{b}$: 
    \setcounter{equation}{25}
	\vspace{-0.3cm} \begin{align}
			&(\mathcal{P}1.4): && \underset{\lfloor\btheta\rceil_b}{\max} ~\underset{k \in \mathcal{K}}{\min} ~ \mbox{SR}_k\left(\wko,\lfloor\thetak\rceil_b\right), ~\text{s.t.} ~\eqref{P1-3}, \label{Ptheta-1} 
		\end{align}
	\begin{figure*}[!t]  
		\begin{alignat}{2}
			&{\sum}_{k=1}^{\mathcal{K}} \mbox{SR}_{k}(\pmw,\lfloor\thetak\rceil_b)&& \geq {\sum}_{k=1}^{\mathcal{K}}q_{k}^{(\iota)}+2{\sum}_{k=1}^{\mathcal{K}}\Re\left\{\left\la \pmb{m}_{k}^{(\iota)},\pmw_k\right\ra\right\}- {\sum}_{k=1}^{\mathcal{K}} \pmw_k^H \pmb{\psi}^{(\iota)}_k \pmw_k,    \label{SSR} \tag{30} \\
			&{\sum}_{k=1}^{\mathcal{K}} \mbox{SR}_k(\wko,\lfloor\btheta\rceil_{b})  &&\geq {q}^{(\iota+1)}+ 2{\sum}_{n=1}^{N}\Re\left\{{\pmb{m}}^{(\iota+1)}(n) e^{j \btheta_n}\right\}. \label{SSRthe} \tag{33}
		\end{alignat} 
		\hrulefill 
	\end{figure*}
	To tackle problem ($\mathcal{P}1.4$) we define 
	\begin{align} \label{convx2}
		\lfloor\theta_n\rceil_b^{(\iota+1),k} = 2\pi - \angle {\pmb{m}}^{(\iota+1)}_k(n), n=1,\dots,N,
	\end{align}
	then we find $\lfloor\thetako\rceil_b$ that maximize the $\mbox{SR}(\wko,\lfloor\btheta\rceil_b)$ as in \eqref{solth}.
	
	The procedure to solve problem ($\mathcal{P}1$) is described in Algorithm \ref{alg1}, which converges to a locally optimal solution of ($\mathcal{P}1$) as formally stated in the following theorem.
	\begin{theorem} \label{theo1}
		The obtained solution by Algorithm \ref{alg1} is a locally optimal solution for problem $(\mathcal{P}1)$.
	\end{theorem}
	\textit{Proof}: See Appendix \ref{AppB}.
	\subsubsection{Complexity Analysis}
    {Algorithm \ref{alg1} addresses problem $(\mathcal{P}1)$ by decoupling the beamforming vector and the IRS PREs in the objective function. The recast problem $(\mathcal{P}1)$ is a Semidefinite Programming (SDP) problem, which can be efficiently solved using the interior-point method \cite{grant2014cvx}. The algorithm’s computational complexity can be characterized in terms of its worst-case runtime and the number of decision variables \cite{labit2002sedumi}. Specifically, in Algorithm \ref{alg1}, the complexity of computing $\pmw$ given $\lfloor\btheta\rceil_{b}$ is $\mathcal{O}(M^{3})$, while that of computing $\lfloor\btheta\rceil_{b}$ given $\pmw$ is $\mathcal{O}(N^{3}(N+1))$. One can notice that the overall complexity grows with the number of IRS PREs, $N$.}
	\vspace{-0.4cm}\subsection{SSR maximization under perfect CSI}
	In this section, we address the problem of maximizing the SSR under the perfect CSI assumption to achieve users' secrecy. 
	The SSR maximization problem can be stated as: \setcounter{equation}{28} 
	\vspace{-0.3cm}\begin{equation} \label{P3}
		(\mathcal{P}2): ~ \underset{\pmw,\lfloor\btheta\rceil_b}{\max}~{\sum}_{k=1}^{\mathcal{K}}\mbox{SR}_k(\pmw,\lfloor\btheta\rceil_b)~\text{s.t.} ~(\ref{P1-2}) \& (\ref{P1-3}).
	\end{equation}    
	Similar to the previous section, we seek $\pmw$ and $\lfloor\btheta\rceil_b$ that maximize the SSR.
	\subsubsection{Sub-Problem for Optimizing the Beamforming Vectors}
	Using (\ref{SRLBR}), the SSR can be expressed as in \eqref{SSR}.
	Therefore, we can rewrite problem ($\mathcal{P}2$) as: \setcounter{equation}{30}
	\vspace{-0.2cm}\begin{align} \label{P4}
		(\mathcal{P}2.1): ~ \underset{\pmw}{\max}  {\sum}_{k=1}^{\mathcal{K}}\mbox{SR}_k(\pmw,\lfloor\thetak\rceil_b), ~ \text{s.t.} ~ (\ref{P1-2}).
	\end{align}
	Problem ($\mathcal{P}2.1$) is a convex problem that can be solved using the Lagrangian multiplier method to generate $\pmw^{(\iota+1)}$ as follows \cite{TTN16}:
	\begin{equation}\label{sr3}
		\wko_k\triangleq\begin{cases}\begin{array}{l}(\pmb{\psi}_k^{(\iota)})^{-1} \pmb{m}_{k}^{(\iota)} ~ \mbox{if}~ \ds {\sum}_{k=1}^{\mathcal{K}}
				||(\pmb{\psi}_k^{(\iota)})^{-1}\pmb{m}_{k}^{(\iota)}||^2\leq P_T,\\
				\left(\pmb{\psi}_k^{(\iota)}+\varpi \pmb{I}_M \right)^{-1} \pmb{m}_{k}^{(\iota)}~ \mbox{otherwise},
			\end{array}
		\end{cases}
	\end{equation} 	
	where bisection method is used to find $\varpi$ such that $\left\|{\pmb{m}_{k}^{(\iota)}}/({\left(\pmb{\psi}_k^{(\iota)}+\varpi \pmb{I}_M\right)})\right\|^2= P_T$.\\
	\subsubsection{Sub-Problem for Optimizing the PREs}
	In this part, we aim to find $\lfloor\thetako\rceil_{b}$ such that it maximizes the SSR. Using (\ref{SRtheta}), we can express the SSR as in \eqref{SSRthe}, where, 
    \vspace{-0.2cm}\begin{align*}
        {q}^{(\iota+1)} \triangleq {\sum}_{k=1}^{\mathcal{K}} {q}_k^{(\iota+1)}, {\pmb{m}}^{(\iota+1)}(n) \triangleq {\sum}_{k=1}^{\mathcal{K}} {\pmb{m}}^{(\iota+1)}_k(n).
    \end{align*}
	We can recast problem ($\mathcal{P}2$) to generate ($\lfloor\thetako\rceil_{b}$) as: \setcounter{equation}{33}
	\vspace{-0.2cm}\begin{subequations} 
		\begin{alignat}{2} 
			&(\mathcal{P}2.2): && ~\underset{\btheta}{\max}  ~ {\sum}_{k=1}^{\mathcal{K}}\mbox{SR}_k(\wko,\lfloor\btheta\rceil_{b}), ~\text{s.t.} ~\eqref{P1-3},
		\end{alignat}
	\end{subequations}
	and the closed-form solution is obtained by, 
	\begin{align} \label{solth2}
		\lfloor\thetako\rceil_{b} =  2\pi - \angle \pmb{m}^{(\iota+1)}(n), n=1,\dots,N.
	\end{align}
	
	Algorithm \ref{alg2} demonstrates the steps to solve problem ($\mathcal{P}2$). Algorithm \ref{alg2} converges to at least a locally optimal solution of ($\mathcal{P}2$) as formally stated in the following theorem.
	\begin{theorem}
		The obtained solution by the AO Algorithm \ref{alg2} is a locally optimal solution for problem $\mathcal{P}2$.
	\end{theorem}
	\textit{Proof}: See Appendix \ref{AppD}.
	\subsubsection{Complexity Analysis}
	The complexity analysis of Algorithm \ref{alg2} is similar to that of Algorithm \ref{alg1}. Hence, the complexity of problem $\mathcal{P}2.1$ and problem $\mathcal{P}2.2$ is about $\mathcal{O} (M^3)$ and $\mathcal{O}(N^{3}(N+1))$, respectively.
	\vspace{-0.3cm}\section{Dealing with Imperfect CSI} \label{sec4}
	\subsection{MAXMIN SR under imperfect CSI}\label{sec3-p2}
	\begin{figure*}[!t] 
		\normalsize 
		\begin{align}
			\mathbf{X}_k &\triangleq \mathbf{\Phi} {\pmb{H}_\text{AR}} \pmw_k (\pmw_k^{(\iota)})^H {\pmb{H}_\text{AR}}^H (\mathbf{\Phi}^{(\iota)})^H+ \mathbf{\Phi}^{(\iota)} {\pmb{H}_\text{AR}} \pmw_k^{(\iota)} \pmw_k^{H} {\pmb{H}_\text{AR}}^H (\mathbf{\Phi}^{(\iota)})^H - \mathbf{\Phi}^{(\iota)} {\pmb{H}_\text{AR}} \pmw_k^{(\iota)} (\pmw_k^{(\iota)})^H {\pmb{H}_\text{AR}}^H (\mathbf{\Phi}^{(\iota)})^H \label{LMI1-1}, \tag{40}\\ 
			& \Delta{\pmb{g}}_{k} \mathbf{X}_k \Delta{\pmb{g}}_{k}^H + 2 \Re \{(\mathbf{x}_k^H + \widehat{h}_{r,k}^H \mathbf{X}_{k}) \Delta{\pmb{g}}_{k}\}+ d_k  \geq  (2^{\varphi_k}-1)\beta_{k}, \forall \|\Delta{\pmb{g}}\|_2 \leq \xi_k, \forall k. \label{Bobx1} \tag{41}
		\end{align}
		\hrulefill 
	\end{figure*}
	In this case, we deal with the IRS-to-users/Eve imperfect CSI. To this end, when only imperfect channel $\widehat{{\pmb{g}}}_{i}$ CSI is available, we adopt the channel modeling in \eqref{ICSIcha}. Nevertheless, in the following analysis, we opt for the case of QPS with PSE, which is the most practical case. One can notice that by setting $b=\infty$ and $\epsilon_R=0$, the PS is reduced to the CPS case. Hence, we can write the minimum SR maximization problem as follows: \setcounter{equation}{35}
	\vspace{-0.3cm}\begin{subequations} \label{P6}
		\begin{alignat}{4} 
			&(\mathcal{P}3): && \underset{\pmw,\lfloor\btheta\rceil_b+\epsilon_R}{\max}   \underset{k}{\min} ~ \widehat{\mbox{SR}}_k(\pmw,\btheta), \label{P2-1} \\
			&~\text{s.t.} && {\sum}_{k = 1}^{\mathcal{K}}\Vert \pmw_k \Vert^2 \leq P_T, \label{P2-2} \\
			& && |e^{(j (\lfloor\btheta\rceil_b+\epsilon_R))}| = 1, \label{P2-3} \\
			& &&\|\Delta{\pmb{g}}_i\|_2 \leq \xi_{i}, i \in \{k,e\}, \forall k. \label{P2-4}
		\end{alignat}
	\end{subequations}
	Problem $(\mathcal{P}3)$ is a nonconvex problem due to the coupling between the beamforming vectors {$\pmw$} and the IRS's PREs {$\lfloor\btheta\rceil_b$} in the objective function {$\eqref{P2-1}$}, and the UMC \eqref{P2-3}. However, problem $(\mathcal{P}3)$ is even more challenging due to the semi-infinite nonconvex constraint \eqref{P2-4} that captures the imperfection of the CSI from the IRS to the users/Eve. In addition to that, the imperfect CSI introduces PE into the IRS's PREs, which makes the problem more challenging. 
	
	To tackle problem $(\mathcal{P}3)$, we first introduce slack variables to decompose the coupling of the beamforming vector and the IRS's PREs in the objective function. To this end, we first substitute \eqref{Cstot2} into \eqref{P2-1}. Then we introduce a slack variable $z$ as the SR's lower bound, ${\varphi}_k$ represents the minimum data rate, and $\mu_{k_e}$ represents the maximum eavesdropping rate of Eve. Thus, problem $(\mathcal{P}3)$ can be recast as:
	\vspace{-0.2cm}\begin{subequations} \label{P6.1}
		\begin{alignat}{2}
			&(\mathcal{P}3.1): &&\underset{\pmw,\lfloor\btheta\rceil_b+\epsilon_R}{\max} ~z, \label{P6-1} \\
			&~\text{s.t.} &&z \leq {\varphi}_k - \mu_{k_e},  \forall k, \label{PwCSI-2x} \\
			& &&{\varphi}_k \leq \widehat{C}_k(\pmw,\lfloor\btheta\rceil_b+\epsilon_R),  \forall \|\Delta{\pmb{g}}_k\|_2 \leq \xi_{k},  \forall k, \label{P6-2} \\
			& &&\mu_{k_e} \geq \widehat{C}_e(\pmw,\lfloor\btheta\rceil_b+\epsilon_R),  \forall \|\Delta{\pmb{g}}_e\|_2 \leq \xi_{e},  \forall k, \label{P6-3} \\
			& &&\eqref{P2-2}, \eqref{P2-3}. \label{P6-4}
		\end{alignat}
	\end{subequations}
	
	One can notice that the constraint \eqref{P2-4} in problem $(\mathcal{P}3)$ has been embedded within the constraints \eqref{P6-2}, and \eqref{P6-3} in problem $(\mathcal{P}3.1)$, since the channel estimation error $\Delta{\pmb{g}}_{i}$ is embedded within the channel definition \eqref{chaCSIdef}. {To this end, we transform the semi-infinite nonconvex constraints \eqref{P6-2} and \eqref{P6-3}, into finite LMIs using the SCA technique and the $\mathcal{S}$-procedure. Then, we tackle the UMC \eqref{P2-3} using the PCCP algorithm.} 
	\subsubsection{Sub-Problem for Optimizing the Beamforming Vectors} \label{subsec3}
	{We start by linearizing the semi-infinite inequalities in \eqref{P6-2}. By substituting \eqref{ckedef2} into \eqref{P6-2}, we can express \eqref{P6-2} as:}
	\begin{align} \label{P3CSI-BnLx2}
		2^{\varphi_k}-1 \leq {\left|({\pmb{g}}_k \pmb{\Phi} {\pmb{H}_\text{AR}})\pmw_{k}\right|^2}/{\left\|({\pmb{g}}_k \pmb{\Phi} {\pmb{H}_\text{AR}})\pmW_{-k}\right\|^2 +\sigk}.
	\end{align}
	{Next, we introduce an auxiliary function $\pmb{\beta} \triangleq [\beta_1, \dots, \beta_{K}]$ to deal with the interference plus noise signal, hence, \eqref{P3CSI-BnLx2} can be expressed as:}
	\begin{subequations}
		\begin{equation} \label{P3CSI-BLx3}
			\left|({\pmb{g}}_k \pmb{\Phi} {\pmb{H}_\text{AR}})\pmw_{k}\right|^2 \geq (2^{\varphi_k}-1)\beta_{k}, \forall k,
		\end{equation}
		\begin{equation} \label{P3CSIn-BLx2}
			\left\|(\pmb{ u}_k \pmb{\Phi} {\pmb{H}_\text{AR}})\pmW_{-k}\right\|^2 +\sigk \leq \beta_{k}, \forall k.
		\end{equation}
	\end{subequations}

    {To address the nonconvex semi-infinite inequalities in \eqref{P3CSI-BLx3}, we substitute their left-hand side with their corresponding lower bounds derived from the following lemma.}
	\begin{mylem} \label{lem1_1}
        At iteration $(\iota)$, let $\wk$ and $\lfloor\thetak\rceil_{b}$ be the optimal solution, then at the point $(\wk,\lfloor\thetak\rceil_{b})$ we can express the linear lower bound of \eqref{P3CSI-BLx3} as: 
        \begin{align} \label{ICSIeqw}
			\left|({\pmb{g}}_k \pmb{\Phi} {\pmb{H}_\text{AR}})\pmw_k\right|^2 \triangleq {\pmb{g}}_k \mathbf{X}_k {\pmb{g}}_k^H, 
		\end{align}
        where $\mathbf{X}_k$, is given in \eqref{LMI1-1}.  
	\end{mylem}
	\Prf Please refer to Appendix \ref{apendixB}. 

    {By substituting \eqref{ICSIeqw} into \eqref{P3CSI-BLx3}, and applying \eqref{ch2-h1} along with Lemma~\ref{lem1_1}, the inequality in \eqref{P3CSI-BLx3} can be recast as in \eqref{Bobx1}, where $d_k \triangleq \widehat{{\pmb{g}}}{k} \mathbf{X}{k} \widehat{{\pmb{g}}}_{k}^H$.}

	
	{Next, we handle the uncertainty of $\{\Delta{\pmb{g}}_k\}$ in \eqref{Bobx1} by employing the $\mathcal{S}$-procedure \cite{boyd1994linear}.}
	\begin{mylem} \label{lem1_2}
		($\mathcal{S}$-procedure) Let $\mathbf{O}_r \in \mathbb{C}^{\mathcal{L} \times \mathcal{L}}$ be a Hermitian matrix, $\mathbf{o}_r \in \mathbb{C}^{\mathcal{L} \times 1}$ be a complex vector, and $\text{o}_r$, for ${r}= 0,\dots,R$ be scalar. A quadratic function of a variable $v$ is defined as: \setcounter{equation}{41}
		\begin{align}
			f_{r}(v) \triangleq v^H \mathbf{O}_{r} x + 2\Re\{\mathbf{o}_{r}^H v\}+{\text{o}}_{r}.
		\end{align}
		The condition $f_{0}(v) \geq 0$ such that $f_{r}(v) \geq 0, ~ r = 1,\dots,R$,  holds, if and only if there exists $\mathfrak{m}_{i} \geq 0, r= 0,\dots,R$, such that,
		\vspace{-0.2cm}\begin{align} 
			\begin{bmatrix}
				\mathbf{O}_{0} & \mathbf{o}_{0} \\ \mathbf{o}_{0}^H & \text{o}_{0}	
			\end{bmatrix} - {\sum}_{r= 0}^{R} \mathfrak{m}_{r} \begin{bmatrix}
				\mathbf{O}_{r} & \mathbf{o}_{r} \\ \mathbf{o}_{r}^H & \text{o}_{r}	
			\end{bmatrix} \succeq 0.
		\end{align} 
	\end{mylem}
	Using Lemma \ref{lem1_2}, equation \eqref{Bobx1} can be transformed into its equivalent LMI as:
	\begin{align} \label{P3CSI-BLX3n2}
		\begin{bmatrix}
			\varpi_{k} \mathbf{I}_{M}+ \mathbf{X}_k & (\widehat{{\pmb{g}}}_{k} \mathbf{X}_k)^H \\ (\widehat{{\pmb{g}}}_{k} \mathbf{X}_k) & d_k - (2^{\varphi_k}-1)\beta_{k} - \eta_{k} \xi_{k}^2	
		\end{bmatrix} \succeq 0, \forall k,
	\end{align}
	where $\pmb{\eta} \triangleq [\eta_{1},\dots,\eta_{K}]^T \geq 0$ are slack variables. 
    
    Although \eqref{P3CSI-BLx3} has been converted into an LMI form, it remains nonconvex due to the nonconvexity nature of the term $2^{\varphi_k} \beta_{k}$. To address this, we employ the SCA technique to approximate the nonconvex constraint \eqref{P3CSI-BLX3n2} with a convex expression. In particular, by applying a first-order Taylor expansion, the term $2^{\varphi_k} \beta_{k}$ can be upper bounded as follows:
	\begin{align} \label{upbeta}
		(\beta_{k}(2^{\varphi_k}))^{ub} \triangleq ((\varphi_k-\varphi_k^{(\iota)})(\beta_{k}^{(\iota)}) \ln2 + \beta_{k})2^{\varphi_k^{(\iota)}},
	\end{align}		
	where $\varphi_k^{(\iota)}, \beta_{k}^{(\iota)}$ are the value of the variables $\varphi_k, \beta_{k}$ at iteration $(\iota)$ in the SCA-based algorithm, respectively. Lastly, substituting \eqref{upbeta} in \eqref{P3CSI-BLX3n2}, the LMIs in \eqref{P3CSI-BLX3n2} can be expressed as:
	\begin{align} \label{P3CSI-BLX3n3}
		\begin{bmatrix}
			\varpi_{k} \mathbf{I}_{M}+ \mathbf{X}_k & (\widehat{{\pmb{g}}}_{k} \mathbf{X}_k)^H \\ (\widehat{{\pmb{g}}}_{k} \mathbf{X}_k) & d_k - (\beta_{k}(2^{\varphi_k}))^{ub}+\beta_{k} - \eta_{k} \xi_{k}^2	
		\end{bmatrix} \succeq 0, \forall k.
	\end{align} 
	Next, to tackle the uncertainties in \eqref{P3CSIn-BLx2}, we leverage the Schur's complement \cite{boyd2004convex}.
	\begin{mylem} \label{lem1_3}
		(Schur's complement) For a Hermitian positive definite matrix $\mathbf{F} \succeq 0$, $\mathbf{D}$, and $\mathbf{S}$, define a Hermitian matrix $\mathbf{A}$ as: 
		\vspace{-0.2cm}\begin{align} 
			\mathbf{A}\triangleq
			\begin{bmatrix}
				\mathbf{S} & \mathbf{D}^H \\ \mathbf{D} & \mathbf{F}
			\end{bmatrix}.
		\end{align} 
	Then, $\mathbf{A}$ is positive semidefinite $(\mathbf{A} \succeq 0)$ if and only if $\Delta{\mathbf{F}} \succeq 0$, and $\mathbf{S}-\mathbf{D}^H\mathbf{F}^{-1}\mathbf{D}\succeq 0$.
	\end{mylem}
	Using Lemma \eqref{lem1_3}, we can equivalently reformulate \eqref{P3CSIn-BLx2} as:
	\begin{align} \label{BobLMIx2n}
		\begin{bmatrix}
			\beta_{k} & \mathbf{t}_k^H \\ \mathbf{t}_{k}^H & \mathbf{I}_{K-1}
		\end{bmatrix} \succeq 0, \forall \|\Delta{\pmb{g}}_k\|_2 \leq \xi_{k}, \forall k,
	\end{align}
	where $\mathbf{t}_{k} \triangleq \left((\widehat{{\pmb{g}}}_{k}^H \mathbf{\Phi} {\pmb{H}_\text{AR}})\mathbf{W}_{-k}\right)^H$, and $\mathbf{W}_{-k} \triangleq [\pmw_1, \dots,\pmw_{k-1},\pmw_{k+1}, \dots,\pmw_{\mathcal{K}}]$. 
	
	Next, to handle \eqref{BobLMIx2n}, we employ the Nemirovski’s Lemma \cite{1369660}.
	\begin{mylem} \label{lem1_4}
		(Nemirovski’s Lemma) Let \(\mathbf{R}\) be a Hermitian matrix, and \(\mathbf{E}\), \(\mathbf{F}\), \(\mathbf{Z}\) be matrices with compatible dimensions. For a scalar \(t > 0\), the following linear matrix inequality holds:
    \begin{align*}
        \mathbf{R} &\succeq \mathbf{E}^H \mathbf{Z} \mathbf{F} + \mathbf{F}^H \mathbf{Z}^H \mathbf{E}, \quad \text{for} \quad \|\mathbf{Z}\| \leq t,
    \end{align*}
    if and only if there exists a scalar \(\alpha \geq 0\) such that:
    \begin{align*}
        \begin{bmatrix}
            \mathbf{R} - \alpha \mathbf{F}^H \mathbf{F} & -t \mathbf{E}^H \\
            -t \mathbf{E} & \alpha \mathbf{I}
        \end{bmatrix} &\succeq 0.
    \end{align*}
	\end{mylem}
	Using Lemma \eqref{lem1_4} and introducing the slack variable  $\pmb{\varkappa} \triangleq [\varkappa_1, \dots, \varkappa_k] \geq 0$, the constraint in \eqref{BobLMIx2n} can be reformulated as:
	\begin{align} \label{P3CSI-BLX4n}
		\begin{bmatrix}
			B_{11,k}& \widehat{t}^H_k & \mathbf{0}_{1 \times M }  \\
			\widehat{t}_k & \mathbf{I}_{K-1} & \xi_{\mathbf{H}_k} (\mathbf{\Phi} {\pmb{H}_\text{AR}} \pmW_{-k})^H  \\  \mathbf{0}_{M \times 1 } & \xi_{\mathbf{H}_k} (\mathbf{\Phi} {\pmb{H}_\text{AR}} \pmW_{-k}) & \varkappa_k \mathbf{I}_{M} 
		\end{bmatrix} \succeq 0, 
	\end{align}
	where 
	\vspace{-0.3cm}\begin{align*}
		B_{11,k} &\triangleq \beta_k - \sigk - \varkappa_k,\\
		\widehat{t}_k &\triangleq \left((\widehat{{\pmb{g}}}_{k} \mathbf{\Phi} {\pmb{H}_\text{AR}})\mathbf{W}_{-k}\right)^H.
	\end{align*}
	Next, we tackle \eqref{P6-3} by firstly substituting \eqref{ckedef2} in  \eqref{P6-3}, and then treating the interference plus noise signal in  \eqref{P6-3} as an auxiliary function $\pmb{\beta}_{e_k}\triangleq [\beta_{e_1},\dots,\beta_{e_k}]$. We hence can recast \eqref{P6-3} as:
	\vspace{-0.3cm}\begin{subequations}
		\begin{alignat}{2}
			&\left|({\pmb{g}}_{e} \mathbf{\Phi} {\pmb{H}_\text{AR}})\pmw_{k}\right|^2 \leq (2^{\mu_{k_e}}-1)\beta_{ke}, &&\forall k, \label{EveLMI1} \\
			&\left\|({\pmb{g}}_{e} \mathbf{\Phi} {\pmb{H}_\text{AR}})\pmW_{-k}\right\|^2 + \sige \geq \beta_{ke}, &&\forall k. \label{EveLMI2}
		\end{alignat}
	\end{subequations}
	{The uncertainties $\{\Delta{\pmb{g}}_e\}$ in constraints \eqref{EveLMI1}, and \eqref{EveLMI2}, can be tackled by a similar approach as in  \eqref{P3CSIn-BLx2}. Hence, the equivalent LMIs of \eqref{EveLMI1}, and \eqref{EveLMI2} are, respectively,}
	\begin{alignat}{2}
		\begin{bmatrix}
			C_{e,k} & \widehat{c}_{ke}^H & \mathbf{0}_{1 \times M} \\
			\widehat{c}_{ke} & 1 & \xi_e (\mathbf{\Phi} {\pmb{H}_\text{AR}} \pmw_k)^H \\
			\mathbf{0}_{M \times 1} & \xi_{e} (\mathbf{\Phi} {\pmb{H}_\text{AR}} \pmw_k) & y_{k} \mathbf{I}_{M}
		\end{bmatrix} &\succeq 0, \label{EveLMI11} \\
		\begin{bmatrix}
			C_{11,k} & -\widehat{c}_{ke}^H & \mathbf{0}_{1 \times M} \\
			-\widehat{c}_{ke} & \mathbf{I}_{K-1} & -\xi_e (\mathbf{\Phi} {\pmb{H}_\text{AR}} \pmW_{-k})^H \\
			\mathbf{0}_{M \times 1} & -\xi_{e} (\mathbf{\Phi} {\pmb{H}_\text{AR}} \pmW_{-k}) & s_{k} \mathbf{I}_{M}
		\end{bmatrix} &\succeq 0, \label{EveLMI22}
	\end{alignat}
	where   $\pmb{y} \triangleq [y_{1}, \dots, y_{K}] \geq 0$ and $\pmb{s} \triangleq [s_{1}, \dots, s_{K}] \geq 0$ are a slack variables, and,
	\vspace{-0.2cm}\begin{align*}
		C_{e,k} & \triangleq (\beta_{ke}(2^{\mu_{k_e}}))^{ub}-\beta_{ke}-\vartheta_{k}, \\
		(\beta_{ke}(2^{\mu_{k_e}}))^{ub} & \triangleq ((\mu_{k_e}-\mu_{k_e}^{(\iota)})(\beta_{ke}^{(\iota)}) \ln2 + \beta_{ke})2^{\mu_{k_e}^{(\iota)}}, \\
		\widehat{c}_{ke} &\triangleq ((\widehat{{\pmb{g}}}_{e}^H \mathbf{\Phi} {\pmb{H}_\text{AR}})\mathbf{W}_{-k})^H, \\
		C_{11,k} &\triangleq \beta_{ke} - \sige - \varrho_{k}.
	\end{align*}
	Eventually, reformulating problem $(\mathcal{P}3.2)$ with \eqref{P3CSI-BLX3n3}, \eqref{P3CSI-BLX4n}, \eqref{EveLMI11}, and \eqref{EveLMI22} yields the following optimization problem: 
	\vspace{-0.1cm}\begin{subequations} \label{P6.3}
		\begin{alignat}{2}
			&(\mathcal{P}3.2):  &&\underset{\pmw}{\max} ~ z, \label{P6.3-1} \\
			&~\text{s.t.}  && z \leq {\varphi}_k - \mu_{k_e}, ~ \forall k, \\
			& && \eqref{P3CSI-BLX3n3}, \eqref{P3CSI-BLX4n}, \eqref{EveLMI11}, \eqref{EveLMI22}, \eqref{P2-2},\\
			& && \pmb{\eta} \geq 0, ~ \pmb{\varkappa} \geq 0, ~ \pmb{y} \geq 0, ~ \pmb{s} \geq 0.  \label{auxvarcon}
		\end{alignat}
	\end{subequations}

    {To this end, we have linearized all the nonlinear constraints \eqref{P6-2}, and \eqref{P6-3}, into their equivalent LMIs \eqref{P3CSI-BLX3n3}, \eqref{P3CSI-BLX4n}, \eqref{EveLMI11}, and \eqref{EveLMI22}; hence, problem ($\mathcal{P}3.2$) can is an SDP problem and can be solved efficiently using standard solvers such as the interior point method, or the CVX toolbox \cite{grant2014cvx}.}
    \begin{algorithm}[t]
\caption{PCCP Algorithm for Solving $(\mathcal{P}3.3)$}
\label{alg3}
\begin{algorithmic}[1]
\State \textbf{Init:} $(\mathbf{w}^{(1)}, \lfloor\boldsymbol{\theta}^{(1)}\rceil_b+\epsilon_R)$, $o_{\max}$, $\nu\ge1$, tolerances $\epsilon_{t_1},\epsilon_{t_2}$, $\iota_{\max}$, set $\iota=1$.
\Repeat
    \State Solve $(\mathcal{P}3.3)$ to obtain $\lfloor\boldsymbol{\theta}_n^{(\iota)}\rceil_b$;
    \If{$\|e^{j\lfloor\boldsymbol{\theta}_n^{(\iota)}\rceil_b} - e^{j\lfloor\boldsymbol{\theta}_n^{(\iota-1)}\rceil_b}\|_1 \le \epsilon_{t_1}$ \ \&\  $Q\le\epsilon_{t_2}$}
        \State \textbf{break};
    \Else
        \State $o^{(\iota+1)} \!=\! \min\{\nu o^{(\iota)},o_{\max}\}$; \ \ $\iota\!\leftarrow\!\iota+1$;
    \EndIf
    \If{$\iota>\iota_{\max}$}
        \State Re-init $\lfloor\boldsymbol{\theta}^{(1)}\rceil_b+\epsilon_R$, set $\nu>1$, $\iota=0$;
    \EndIf
\Until{convergence}
\State \textbf{Output:} $\boldsymbol{\theta}^*=\lfloor\boldsymbol{\theta}^{(\iota)}\rceil_b+\epsilon_R$.
\end{algorithmic}
\end{algorithm}
	\subsubsection{Sub-Problem for Optimizing the PREs} \label{subsec4}
	{At fixed $\pmw$, we aim to find the next feasible point $\lfloor \thetako \rceil_b$, such that, $\mbox{SR}_k(\wko_k,\lfloor \thetako \rceil_b) > \mbox{SR}_k(\wko_k,\lfloor\thetak\rceil_{b})$. Similar to the previous section, constraints $\eqref{P6-2}$ and $\eqref{P6-3}$ can be recast as their equivalent LMIs \eqref{P3CSI-BLX3n3}, \eqref{P3CSI-BLX4n}, \eqref{EveLMI11}, and \eqref{EveLMI22}, respectively.}
    
	{However, Problem $(\mathcal{P}3)$ still nonconvex due to the noncovixty nature of the UMC \eqref{P2-3}. To this end, we leverage the PCCP method to tackle the UMC \eqref{P2-3}. The idea of the PCCP method is to introduce slack variables that relax the objective function, allowing violations of the UMC to be penalized through the sum of these violations. The solution obtained upon convergence of the PCCP method is an approximate first-order optimal solution to the original problem \cite{9525400}.}
	
	{To implement the PCCP method, we introduce an auxiliary variable set $\mathbf{\mathcal{Q}} \triangleq \{\mathcal{Q}_{n}|n \in N\}$, where each $\mathcal{Q}_{n}$ satisfies $\mathcal{Q}_n \triangleq |e^{(j \lfloor \btheta_n \rceil_b)}|^* |e^{(j \lfloor \btheta_n \rceil_b)}|$. Thus, the UMC \eqref{P2-3} can be written as $\mathcal{Q}_n \leq |e^{(j \lfloor \btheta_n \rceil_b)}|^* |e^{(j \lfloor \btheta_n \rceil_b)}| \leq \mathcal{Q}_n $. The nonconvex inequality $\mathcal{Q}_n \leq |e^{(j \lfloor \btheta_n \rceil_b)}|^* |e^{(j \lfloor \btheta_n \rceil_b)}|$ is then approximated using its first order Taylor expansion as $\mathcal{Q}_n \leq 2 \Re \{|e^{(j \lfloor \btheta_n \rceil_b)}|^* |e^{(j \lfloor \thetak_b \rceil_b)}| - |e^{(j \lfloor \thetak_b \rceil_b)}|^* |e^{(j \lfloor \thetak_b \rceil_b)}|\}$. By applying the PCCP framework, we penalize the objective function \eqref{P6-1}, and thus reformulate problem $(\mathcal{P}3.1)$ as:}
	\vspace{-0.1cm}\begin{subequations} \label{P3CSIthBL}
		\begin{alignat}{2}
			&(\mathcal{P}3.3):  &&\underset{\btheta}{\max} ~ z - o \mathcal{T}, \label{P3CSIthBL-1} \\
			&~\text{s.t.}  && \eqref{P3CSI-BLX3n3}, \eqref{P3CSI-BLX4n}, \eqref{EveLMI11}, \eqref{EveLMI22}, \label{P3CSIthBL-constraints} \\
			& && |e^{(j \lfloor \thetak_n \rceil_b)}|^* |e^{(j \lfloor \thetak_n \rceil_b)}| \leq \mathcal{Q}_n + d_n, \label{P3CSIthBL-BLx1} \\
			& && \mathcal{Q}_n - \hat{d}_n \leq 2 \Re \left\{|e^{(j \lfloor \thetak_n \rceil_b)}|^* |e^{(j \lfloor \thetak_n \rceil_b}|\right\} \nonumber \\ 
			& && - |e^{(j \lfloor\thetak_n\rceil_{b})}|^* |e^{(j \lfloor\thetak_n\rceil)}|, \label{P3CSIthBL-BLx2} \\
			& && \mathcal{Q}_n \geq 0, ~ \forall n \in N, \label{P3CSIthBL-BLx3}
		\end{alignat}
	\end{subequations}
	where $\mathcal{T} \triangleq {\sum}_{n = 1}^N d_n+\hat{d}_n$ is the penalty term, $\mathbf{d} \triangleq \{d_n,\hat{d}_n\}$ denotes the slack variable introduced to relax the UMC \eqref{P2-3}, and $o$ is the regularization factor. The regularization factor $o$ is imposed to scale the penalty term $\mathcal{T}$ to control the UMC \eqref{P2-3}.
	\begin{algorithm}
		\caption{Proposed AO Algorithm for Solving Problem $(\mathcal{P}3)$} \label{alg4}
		\begin{algorithmic}[1]
			\State \textbf{Initialize:} $(\pmb{\text{w}}^{(1)},\lfloor \btheta^{(1)} \rceil_{b} + \epsilon_R$, $\iota=1$, and convergence tolerance $\epsilon_t > 0$.
			\State \textbf{Repeat} 
			\State Update $\wko$ by solving $(\mathcal{P}3.2)$, and $\lfloor \thetako_n \rceil_b$ by solving problem $(\mathcal{P}3.2)$;
			\State \textbf{if} $\frac{\left|\underset{k}{\min} \mbox{SR}_{k}(\wko,\lfloor \thetako_n \rceil_b+\epsilon_R)-\underset{k}{\min} \mbox{SR}_{k}(\wk,\lfloor\thetak\rceil_b+\epsilon_R)\right|}{\underset{k}{\min} \mbox{SR}_{k}(\wk,\lfloor\thetak\rceil_b+\epsilon_R)} \leq \epsilon_t$.
			\State \textbf{Then} $\lfloor \thetak \rceil_b \leftarrow \lfloor \thetako_n \rceil_b$, $\wk\leftarrow \wko$ and terminate. 
			\State \textbf{Otherwise} $\iota\leftarrow \iota+1$ and continue.
			\State \textbf{Output} $(\wk_k, \lfloor\thetak\rceil_b+\epsilon_R)$.
		\end{algorithmic}
	\end{algorithm}
	Problem $(\mathcal{P}3.3)$ is an SDP problem that the CVX toolbox can solve. Unlike the semi-definite relaxation (SDR) method, which is the conventional method to tackle UMC, the PCCP method is guaranteed to find a feasible point for problem $(\mathcal{P}3.3)$ \cite{9774882}. The PCCP algorithm applied to solve problem $(\mathcal{P}3.3)$ is outlined in Algorithm \ref{alg3}. 
	The following points are emphasized from Algorithm \ref{alg3}:
	\begin{enumerate}[label=(\alph*)]
		\item We invoke $o_{\text{max}}$  to prevent numerical instability in the algorithm when the penalty parameter $o$ grows too large \cite{9505311};
		\item The stopping criteria $Q \leq \epsilon_{t_2}$ ensures satisfaction of the UMC \eqref{P2-3}, provided that the tolerance $\epsilon_{t_2}$ is small \cite{9180053};
		\item The convergence of Algorithm \ref{alg3} is controlled by $\left\|e^{(j \lfloor \thetako\rceil_b)}-e^{(j \lfloor \thetak\rceil_b)}\right\|_1 \leq \epsilon_{t_1}$, which controls the incremental change in the PREs PS updates.
	\end{enumerate}
	Algorithm \ref{alg4} demonstrates the AO algorithm to solve problem ($\mathcal{P}3$), which converges to at least a locally optimal solution of ($\mathcal{P}3$), and can be proved in the following theorem.
	\begin{theorem} \label{theo2}
		The AO Algorithm \ref{alg4} generates a locally optimal solution for problem $(\mathcal{P}3)$.
	\end{theorem}
	\textit{Proof}: See Appendix \ref{AppD}.
	\subsubsection{Complexity Analysis}
    Problems $(\mathcal{P}3.2)$ and $(\mathcal{P}3.3)$ are convex problems, and involve linear constraints and LMIs $\eqref{P3CSI-BLX3n3}, \eqref{P3CSI-BLX4n}, \eqref{EveLMI11}, \eqref{EveLMI22}$, hence, they can be solved efficiently using the interior point method \cite{grant2014cvx}. The complexity of Algorithm \ref{alg4} is determined by its worst-case runtime and the number of operations performed \cite{9180053}. For problem $(\mathcal{P}3.3)$, the number of variables is $c\triangleq 2M$, the size of LMI \eqref{P3CSI-BLX3n3} is $a_1\triangleq MN+M+1$, the size of LMIs \eqref{P3CSI-BLX4n}, and \eqref{EveLMI22} is $a_2\triangleq 2M+1$, and the size of  the LMI \eqref{EveLMI11} is $a_3\triangleq 2M+1$. Thus, the complexity of solving problems $(\mathcal{P}3.2)$ and $(\mathcal{P}3.3)$, respectively,
	\begin{align}
		\mathcal{O}_{\pmw} &\triangleq \mathcal{O}\left\{\sqrt{b_1+2} (c)(c^2+c b_2+b_3+(c+1)^2)\right\}, \label{comx1} \\ 
		\mathcal{O}_{\pmb{\Phi}} &\triangleq \mathcal{O}\left\{\sqrt{b_1+4N} (2N)(4N^2+2N b_2+b_3+4NM)\right\}, \label{comx2}
	\end{align}	
	where,
	\begin{align*}
		{b_1} & \triangleq {\sum}_{k}^{\mathcal{K}}({a_1}+{a_2})+2k({a_2}+{a_3})+(k-2)({a_1}+{a_3}), \\
		{b_2} & \triangleq {\sum}_{k}^{\mathcal{K}}({a_1}^2+{a_2}^2)+2k({a_2}^2+{a_3}^2)+(k-2)({a_1}^2+{a_3}^2), \\
		{b_3}&\triangleq{\sum}_{k}^{\mathcal{K}}({a_1}^3+{a_2}^3)+2k({a_2}^3+{a_3}^3)+(k-2)({a_1}^3+{a_3}^3).
	\end{align*} 
	\subsection{SSR maximization under imperfect CSI}
	Previous research, such as \cite{10256584}, has considered a similar problem with the presence of a direct link between Alice and the users and Eve.  
	The SSR maximization problem under imperfect CSI from the IRS to users/Eve can be formally formulated as:
	\vspace{-0.3cm}\begin{algorithm}
		\caption{Proposed AO Algorithm for Solving Problem $(\mathcal{P}4)$} \label{alg6}
		\begin{algorithmic}[1]
			\State \textbf{Initialize:} $(\pmb{\text{w}}^{(1)}, \lfloor\btheta^{(1)}\rceil_b+\epsilon_R)$, convergence tolerance $\epsilon_t > 0$,  and Set $\iota=1$. 
			\State \textbf{Repeat} 
			\State Update $\wko$ by solving problem $\mathcal{P}4.2$, and $\lfloor \thetako_n \rceil_b+\epsilon_R$ by solving problem $\mathcal{P}4.3$;
			\State \textbf{if} $\frac{\left|\sum\limits_{k=1}^{\mathcal{K}} \mbox{SR}_{k}(\wko,\lfloor \thetako_n \rceil_b+\epsilon_R)-\sum\limits_{k=1}^{\mathcal{K}} \mbox{SR}_{k}(\wk,\lfloor\thetak\rceil_b+\epsilon_R)\right|}{\sum\limits_{k=1}^{\mathcal{K}} \mbox{SR}_{k}(\wk,\lfloor\thetak\rceil_b+\epsilon_R)} \leq \epsilon_t$.
			\State \textbf{Then} $\lfloor\thetak\rceil_{b}\leftarrow \lfloor \thetako_n \rceil_b$, $\wk\leftarrow \wko$ and terminate. 
			\State \textbf{Otherwise} $\iota\leftarrow \iota+1$ and continue.
			\State \textbf{Output} $(\wk_k, \lfloor\thetak\rceil_b+\epsilon_R)$.
		\end{algorithmic}
	\end{algorithm}	
	\begin{subequations} \label{P9}
		\begin{alignat}{2}
			&(\mathcal{P}4):  &&\underset{\pmw,\btheta}{\max} ~ {\sum}_{k=1}^{\mathcal{K}} \widehat{\mbox{SR}}_{k}(\pmw,\btheta), \label{P9-1} \\
			&~\text{s.t.}  && \eqref{P2-2}, \eqref{P2-3}, \eqref{P2-4}.
		\end{alignat}
	\end{subequations}
	Similar to $(\mathcal{P}3)$, problem $(\mathcal{P}4)$ is a nonconvex problem due to the coupling between the beamforming vector and the IRS's PREs within the objective function, and the UMC \eqref{P2-3}. 
	
	To tackle problem $(\mathcal{P}4)$, we first substitute \eqref{Cstot2} in \eqref{P9-1}, and introduce the slack variables $\tilde{z}$ as lower bound of the SSR, $\tilde{\varphi}_k$ and $\tilde{\mu}_{k_e}$  as the minimum data rate, and the maximum eavesdropping rate of Eve, respectively, hence, problem $(\mathcal{P}4)$ can be recast as:
	\vspace{-0.3cm}\begin{subequations} \label{PwCSI4}
		\begin{alignat}{2}
			&(\mathcal{P}4.1):  &&\underset{\pmw, \lfloor\btheta\rceil_b}{\max} ~ \tilde{z}, \label{PwCSI4-1} \\
			&~\text{s.t.}  && \tilde{z} \leq {\sum}_{k=1}^{\mathcal{K}} \left(\tilde{\varphi}_k - \tilde{\mu}_{e_k}\right), \forall k, \label{PwCSI4-2x} \\
			& && \tilde{\varphi}_k \leq \widehat{C}_k(\pmw,\lfloor\btheta\rceil_b+\epsilon_R), \forall \|\Delta{\pmb{g}}_k\|_2 \leq \xi_{k}, \label{PwCSI4-2} \\
			& && \tilde{\mu}_{k_e} \geq \widehat{C}_e(\pmw,\lfloor\btheta\rceil_b+\epsilon_R), \forall \|\Delta{\pmb{g}}_e\|_2 \leq \xi_{e},   \label{PwCSI4-3} \\
			& && \eqref{P2-2}, \eqref{P2-3}.		\end{alignat}
	\end{subequations}
	To tackle problem $(\mathcal{P}4.1)$, we use a similar approach that dealt with problem $(\mathcal{P}3.1)$.
	\subsubsection{Sub-Problem for Optimizing the Beamforming Vectors}	
	Similar to the analysis to linearize $\eqref{P3CSI-BLx3}$ and $\eqref{P3CSIn-BLx2}$ in section \ref{sec4}. First, we introduce an auxiliary variable $\tilde{z}$ as the lower bound of the SSR. Then, we express $\eqref{PwCSI4-2}$ and $\eqref{PwCSI4-3}$ with their equivalent LMIs \eqref{P3CSI-BLX3n3}, \eqref{P3CSI-BLX4n}, \eqref{EveLMI11} and \eqref{EveLMI22}, respectively. Thus, problem $(\mathcal{P}4.1)$ can be recast as: 
	\vspace{-0.1cm}\begin{subequations} \label{P9.3}
		\begin{alignat}{2}
			&(\mathcal{P}4.2):  &&\underset{\pmw, \btheta}{\max} ~ \tilde{z}, \label{PwCSI4-3-1} \\
			&~\text{s.t.}  && \tilde{z} \leq {\sum}_{k=1}^{\mathcal{K}} \left(\tilde{\varphi}_k - \tilde{\mu}_{e_k}\right),  \forall k, \label{PwCSI4-3-2x} \\
			& && \eqref{P2-2},\eqref{P3CSI-BLX3n3}, \eqref{P3CSI-BLX4n}, \eqref{EveLMI11}, \eqref{EveLMI22}, \eqref{auxvarcon}.
		\end{alignat}
	\end{subequations}
	Problem ($\mathcal{P}4.2$) is a convex problem that can be solved using the CVX toolbox. 
	\subsubsection{Sub-Problem for Optimizing the PREs} \label{subsec4SSR}
	Similarly, we can express $\eqref{PwCSI4-2}$ and $\eqref{PwCSI4-3}$ with their equivalent LMIs \eqref{P3CSI-BLX3n3}, \eqref{P3CSI-BLX4n}, \eqref{EveLMI11} and \eqref{EveLMI22}, respectively.
	To tackle the UMC \eqref{P2-3}, we use the PCCP method as described in Algorithm \ref{alg3}.
	Finally, the steps to solve problem ($\mathcal{P}4$) are described in Algorithm \ref{alg6}. 
	\begin{theorem}
		The obtained solution by Algorithm \ref{alg6} is a locally optimal solution for problem $\mathcal{P}4$.
	\end{theorem}
	\textit{Proof}: See Appendix \ref{AppD}.
	\subsubsection{Complexity Analysis}
	The complexity analysis of Algorithm \ref{alg6} is similar to  Algorithm \ref{alg4}. Hence, the complexity of problem $\mathcal{P}4$ is expressed in \eqref{comx1} and \eqref{comx2}.
	\vspace{-0.3cm}\section{Dealing with Eve's Unknown CSI} \label{secAN2}
			\begin{algorithm}
		\caption{Proposed AO Algorithm for Solving Problem $(\mathcal{P}5)$} \label{alg7}
		\begin{algorithmic}[1]
			\State \textbf{Initialize:} $(\pmb{\text{w}}^{(1)}, \lfloor\btheta^{(1)}\rceil_{b}+\epsilon_R)$, convergence tolerance $\epsilon_t > 0$,  and Set $\iota=1$. 
			\State \textbf{Repeat} 
			\State Update $\wko$ by solving problem $(\mathcal{P}5.2)$, and $\lfloor \thetako_n \rceil_b$ by solving problem $(\mathcal{P}5.3)$;
			\State \textbf{if} $\frac{\left|{\sum}_{k=1}^{\mathcal{K}} \|\wko\|^2-{\sum}_{k=1}^{\mathcal{K}} \|\wk\|^2\right|}{{\sum_{k}^{\mathcal{K}}}\|\wk\|^2} \leq \epsilon_t$.
			\State \textbf{Then} $\lfloor\thetak\rceil_{b}\leftarrow \lfloor \thetako_n \rceil_b$, $\wk\leftarrow \wko$ and terminate. 
			\State \textbf{Otherwise} $\iota\leftarrow \iota+1$ and continue.
			\State \textbf{Output} $(\wk_k, \lfloor\thetak\rceil_{b}+\epsilon_R)$.
		\end{algorithmic}
	\end{algorithm}
	In this case, we assume that Alice can't obtain Eve's CSI (even just the imperfect CSI).  
	To provide secure communications for \emph{all users}, we propose to optimize a minimum transmit power $P_T$ subject to a QoS constraint (i.e. $\gamma_k \ge \gamma_e$) to allow us to use the residual power available at Alice as AN so as to decrease the SINR at Eve $\gamma_{e}$. Unlike other research  \cite{9146177,9764813}, that only deals with a single user and assumes perfect CSI under IRS' PREs CPS, we address the power minimization problem under the multi-users settings, where only partial users' CSI is available, and the IRS's PREs are modeled by the QPS and suffer PSE due to the imperfect CSI. To this end, we can express the total transmit power $P_{tot}$ as follows:
	\vspace{-0.3cm}\begin{align} \label{respo}
		P_{tot} \triangleq P_{T}+P_{J},
	\end{align}
	where the residual power $P_J\triangleq\sum_{k=1}^{\mathcal{K}}\mathbf{E}\{\|\pmb{\mathbf{Vr}}\|^2\}$ is used as AN, and $\mathbf{Vr}$ is the AN vector, while $\mathbf{V} \in \mathbb{C}^{M \times M-1}$ is the null space of the user's channel with $\pmb{h}_{k}(\lfloor\btheta\rceil_{b}+\epsilon_R) \mathbf{V} = 0$, and $\mathbf{r} \in \mathbb{C}^{M-1}$ is a random vector whose elements are independent Gaussian random variables with mean $0$ and variance $\sigma_r \triangleq \frac{P_J}{M-1}$. Hence, we can express the SINR at user-$k$ as in \eqref{SINRICSI}, and Eve's eavesdropping rate as:
	\begin{align} \label{SINRAN}
		{\gamma}_e(\pmw,\lfloor\btheta\rceil_b+\epsilon_R)&\triangleq\frac{|{\widehat{\pmb{h}}}_{e}(\lfloor\btheta\rceil_b+\epsilon_R)\pmw_k|^2}{{\widehat{\rho}}_e+|\widehat{\pmb{h}}_{e}(\lfloor\btheta\rceil_b+\epsilon_R)\mathbf{Vr}|^2}.
	\end{align}
	To that end, we can formulate the problem as:
	\begin{subequations} \label{PAB}
		\begin{alignat}{2}
			&(\mathcal{P}5):  &&\underset{\pmw,\btheta}{\min} ~ P_T, \label{PAB-1} \\
			&~\text{s.t.}  &&\gamma_k(\pmw,\lfloor\btheta\rceil_b+\epsilon_R) \geq \gamma \sigk, \label{PAB-4} \\
			& &&\eqref{P2-2}, \eqref{P2-3}, \eqref{P2-4}. \label{PAB-2}
		\end{alignat}
	\end{subequations}
	where $\gamma$ represents the QoS constraint at user-$k$. Due to the coupling of $\pmw$ and $\btheta$ in \eqref{PAB-4}, problem $(\mathcal{P}3)$ is a nonconvex problem. 
	\subsubsection{Sub-Problem for Optimizing the Beamforming Vectors} \label{subsec1-AN}
	similar to the previous sections, we propose an AO method to tackle problem $(\mathcal{P}5)$, by designing the beamform vector $\pmw$ with fixed $\btheta$. Hence, problem $(\mathcal{P}5)$ can be recast as:
	\begin{subequations} \label{PAB02}
		\begin{alignat}{2}
			&(\mathcal{P}5.1):  &&\underset{\pmw}{\min} ~ P_T, \label{PAB02-1} \\
			&~\text{s.t.}  && \left|({\pmb{g}}_k \pmb{\Phi} {\pmb{H}_\text{AR}})\pmw_{k}\right|^2 \geq (2^{\gamma_k}-1)\hat{\beta}_{k},  \forall k, \label{PAB02-3} \\
			& && \left\|({\pmb{g}}_k \pmb{\Phi} {\pmb{H}_\text{AR}})\pmW_{-k}\right\|^2 + \sigk \leq \hat{\beta}_{k}, \forall k, \label{PAB02-4} \\
			& &&\eqref{P2-2}, {\eqref{P2-4}}. \label{PAB02-5}
		\end{alignat}
	\end{subequations}
	where $\hat{\pmb{\beta}} \triangleq [\hat{\beta}_1, \dots, \hat{\beta}_{K}]$ is an auxiliary function. Similar to the analysis to linearize $\eqref{P3CSI-BLx3}$ and $\eqref{P3CSIn-BLx2}$ in section \ref{sec4}, we can express $\eqref{PAB02-3}$ and $\eqref{PAB02-4}$ with their equivalent LMIs \eqref{P3CSI-BLX3n3}, \eqref{P3CSI-BLX4n}, \eqref{EveLMI11} and \eqref{EveLMI22}, respectively. Thus, problem $(\mathcal{P}5.1)$ can be recast as: 
	\vspace{-0.1cm}\begin{subequations} \label{PAB03}
		\begin{alignat}{2}
			&(\mathcal{P}5.2):  &&\underset{\pmw}{\min} ~ P_T, \label{PAB03-1} \\
			&~\text{s.t.}  && \eqref{P2-2}, {\eqref{P2-4}},\eqref{P3CSI-BLX3n3}, \eqref{P3CSI-BLX4n}, \eqref{EveLMI11}, \eqref{EveLMI22}, \eqref{auxvarcon}.
		\end{alignat}
	\end{subequations}
	Problem ($\mathcal{P}5.2$) is a convex problem that can be solved using the CVX toolbox. 
	\subsubsection{Sub-Problem for Optimizing the PREs}
	Similarly, we aim to find $\lfloor\thetako\rceil_{b}$ that minimizes $P_T$ while meeting the QoS constraint \eqref{PAB-4}. This problem is a search problem to find any feasible $\lfloor\btheta\rceil_{b}$ satisfying the QoS \eqref{PAB-4} and the UMC constraint \eqref{P2-3}, where the obtained result can be the optimal solution. Hence, one can use the residual power \eqref{respo} and the PCCP algorithm as described in Algorithm \ref{alg3} to find the optimal $\lfloor\btheta\rceil_{b}$. 
	
	Finally, the algorithm to solve problem ($\mathcal{P}5.3$) is described in Algorithm \ref{alg7}. 
	\begin{theorem}
		The solution obtained using the AO Algorithm~\ref{alg7} is a locally optimal solution to problem $(\mathcal{P}5)$.
	\end{theorem}
	\textit{Proof}: See Appendix \ref{AppD}.
	\subsubsection{Complexity Analysis} Algorithm  \ref{alg7} involves linear constraints and LMIs $\eqref{P3CSI-BLX3n3}, \eqref{P3CSI-BLX4n}, \eqref{EveLMI11}, \eqref{EveLMI22}$, hence, its complexity is similar to Algorithm \ref{alg4}, which can be expressed as in \eqref{comx1} and \eqref{comx2}. 
	\vspace{-0.5cm}\section{Simulations Results} \label{sec5}
	\begin{table}
		\centering
		\caption{Numerical Parameters}
		\begin{tabularx}{0.5\textwidth} {
				| >{\raggedright\arraybackslash}X
				| >{\centering\arraybackslash}X
				| >{\raggedleft\arraybackslash}X | }
			\hline
			Parameter & Numerical Value \\
			\hline
			Noise power density ($\sigma_i$) & $-174$ dBm/Hz\\
			\hline
			Alice Transmission power ($P_T$) & $20$ dBm\\
			\hline
			Antennas' Gain $(G_{\text{\text{A}}})$ and $(G_{\text{IRS}})$ & $5$ dBi \\
			\hline
			PS bits  $(b)$ & $3$\\
			\hline
			the convergence tolerance $\epsilon_t$ & $10^{-3}$  \\
			\hline
			simulation initial settings & $\beta_{k}^{(\iota)}= 1$, $\beta_{ke}^{(\iota)}= 1$, $o^{(1)}=10$, $o_{\text{max}}=30$, $\delta_{k}=\delta_{e}=0.02$.\\
			\hline
		\end{tabularx}
		\label{T2}
	\end{table}
	To evaluate the performance of our proposed algorithms, we conduct extensive simulations using MATLAB and CVX toolboxes. In this setup, we consider the downlink transmission, where Alice is positioned at $(15,0,15)$, and the IRS is placed at $(0,25,40)$. Legitimate users are randomly distributed within a $(60m \times 60m)$ area located to the left of Alice. The Eve is randomly located in $(60m \times 60m)$ outside the user's area. Notably, when Eve is positioned in close proximity to a legitimate user, ensuring a positive secrecy is unattainable~\cite{6772207}. In such cases, alternative techniques, such as cryptographic encryption or friendly jamming, can be employed to enhance secrecy. 
	
    The direct path loss factor of Alice-to-IRS is modelled as ${\beta_{_{\text{{AR}}}}} \triangleq G_{\text{\text{A}}}+G_{\text{IRS}}-35.9-22 \log_{10}(d_{_{\text{{AR}}}})$ in dB, where $d_{_{\text{{AR}}}}$ is the distance between Alice and the IRS in meters, and $G_{\text{IRS}}$ is the IRS elements' antenna gain \cite{BOL19}. The IRS-users/Eve is ${\beta_{\text{R}i}} \triangleq G_{\text{IRS}}-33.05-30 \log_{10}(d_{\text{R}i})$ dB, for $i\in\{k,e\}$, $d_{\text{R}i}$ is the distance between the IRS and the users and Eve in meters. 
	The small-scale fading channel gain $\widehat{{\pmb{g}}}_{i}$ for $ i \in \{k,e\}$ is modeled as a Rician fading channel with K-factor$=3$. The spatial correlation matrix is $[R_{Ri}]_{q,\bar{q}} \triangleq \exp{(j\pi (q-\bar{q}) \sin \hat{\vartheta} \sin \hat{\aleph})}$ for $i\in \{k,e\}$, where $\hat{\aleph}$ is the elevation angle and $\hat{\vartheta}$ is the azimuth angle. The elements of the Alice-to-IRS channel are generated by $[G_{_{\text{{AR}}}}]_{a,b} \triangleq \exp({j\pi \left((b-1) \sin\overline{\Theta}_b \sin \overline{\vartheta}_b+(a-1) \sin(\Theta_n) \sin(\vartheta_b)\right)})$, where ${\Theta}_n \in (0,2 \pi)$, ${\vartheta}_n \in (0,2 \pi)$, and $\overline{\Theta}_n \triangleq \pi-\Theta_n$, and $\overline{\vartheta}_n \triangleq \pi+\vartheta_n$ \cite{kammoun2020asymptotic}.   

    In the imperfect CSI case, the channel estimation error bounds are defined as $\xi_i \triangleq \delta_i \|\widehat{{\pmb{g}}}_i\|_2, \forall k$, where $\delta_i \in [0,1), i\in \{k,e\}$ is the relative amount of CSI uncertainty. The special case $\delta = 0$,  corresponds to perfect CSI at Alice for the IRS-to-user/Eve reflected channels. Unless stated otherwise, the simulation parameters are defined in Table \ref{T2}. Lastly, the computed rates are multiplied by $\log_2(e)$ to convert them to bps/Hz.
	\begin{figure}[t!]
		\centering
		\includegraphics[width=0.33\textwidth]{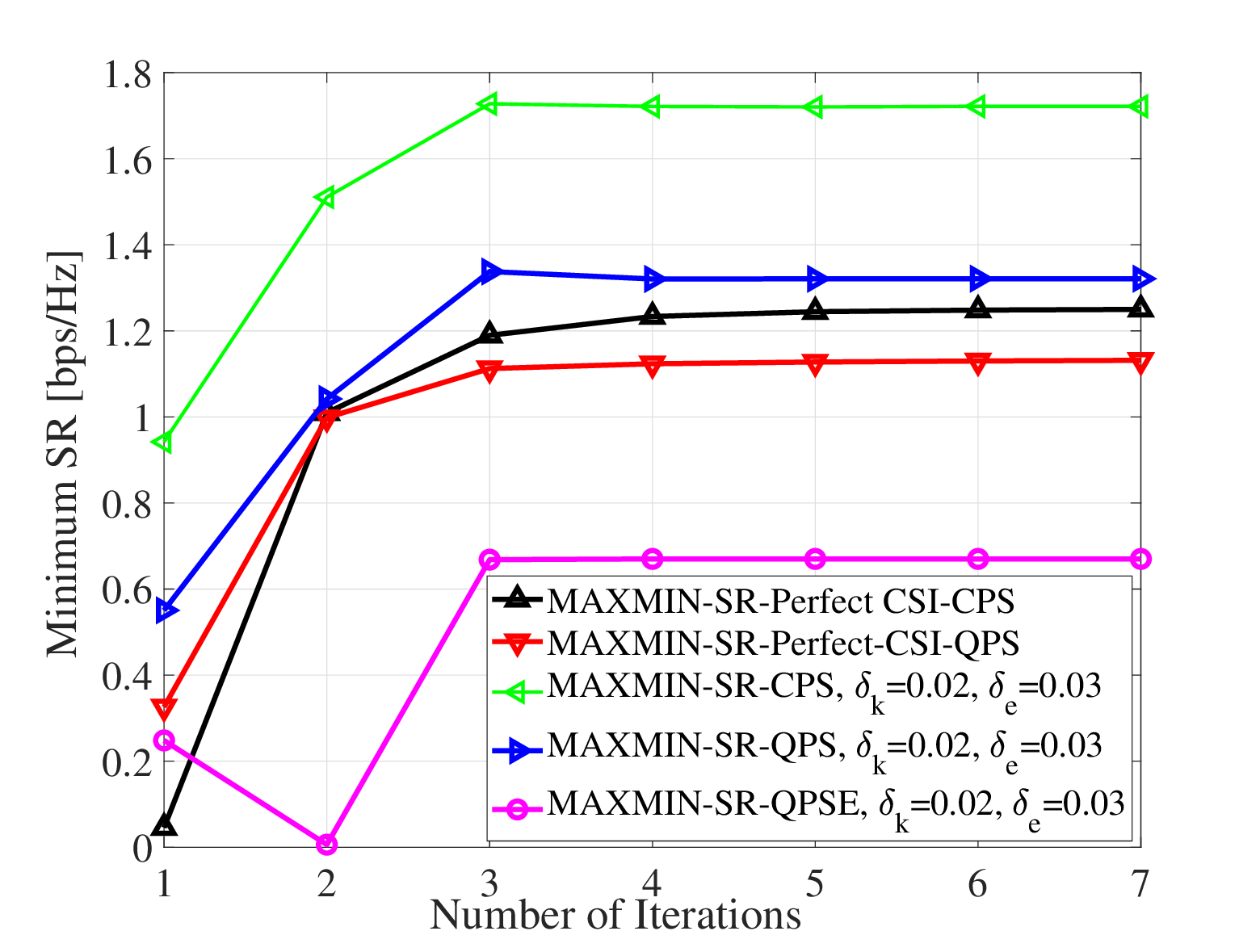}
		\caption{Convergence rate of the Max-Min algorithm with $M=10$, $\mathcal{K}=6$, $N=16$.}
		\label{Fig2x}
	\end{figure}
	\begin{figure}[t!]
		\centering
		\includegraphics[width=0.33\textwidth]{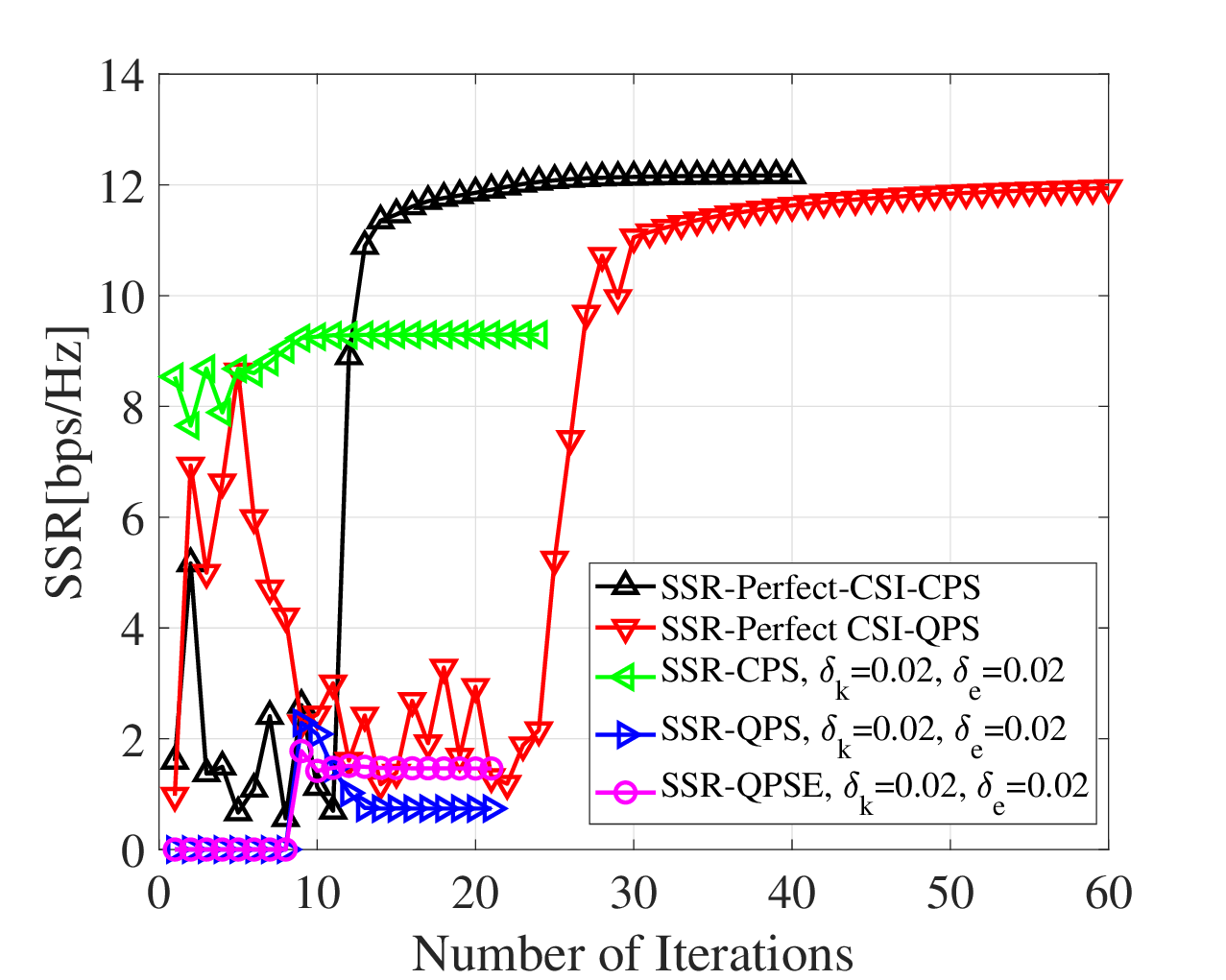}
		\caption{Convergence rate of SSR maximization with $M=10$, $\mathcal{K}=6$, $N=16$.}
		\label{Fig3} 
	\end{figure}
	
	Fig. \ref{Fig2x}  illustrates the convergence rate of the Max-Min algorithms under the perfect and imperfect CSI for the CPS, the QPS, and the QPSE cases with $M=10$, $\mathcal{K}=6$, and $N=16$.  All algorithms achieve convergence within a few iterations. The obtained results show the robustness of the proposed algorithm, even when the IRS's PREs PS are subject to phase error and CSI uncertainty, the system still can achieve secrecy for \emph{all users}. The SSR counterpart convergence rate is shown in Fig. \ref{Fig3}. One can notice the oscillation at the beginning of the convergence curve at the first iterations before the convergence, which occurs when the SSR algorithm is trying to provide secrecy to \emph{all users}. It can be noticed that the convergence is unreachable until the algorithm drops the users with zero secrecy, which is visible in all cases. The results show that the SSR algorithm suffers from the PS error more than the Max-Min algorithm.
	
\begin{figure}[t!]
		\centering
		\includegraphics[width=0.33\textwidth]{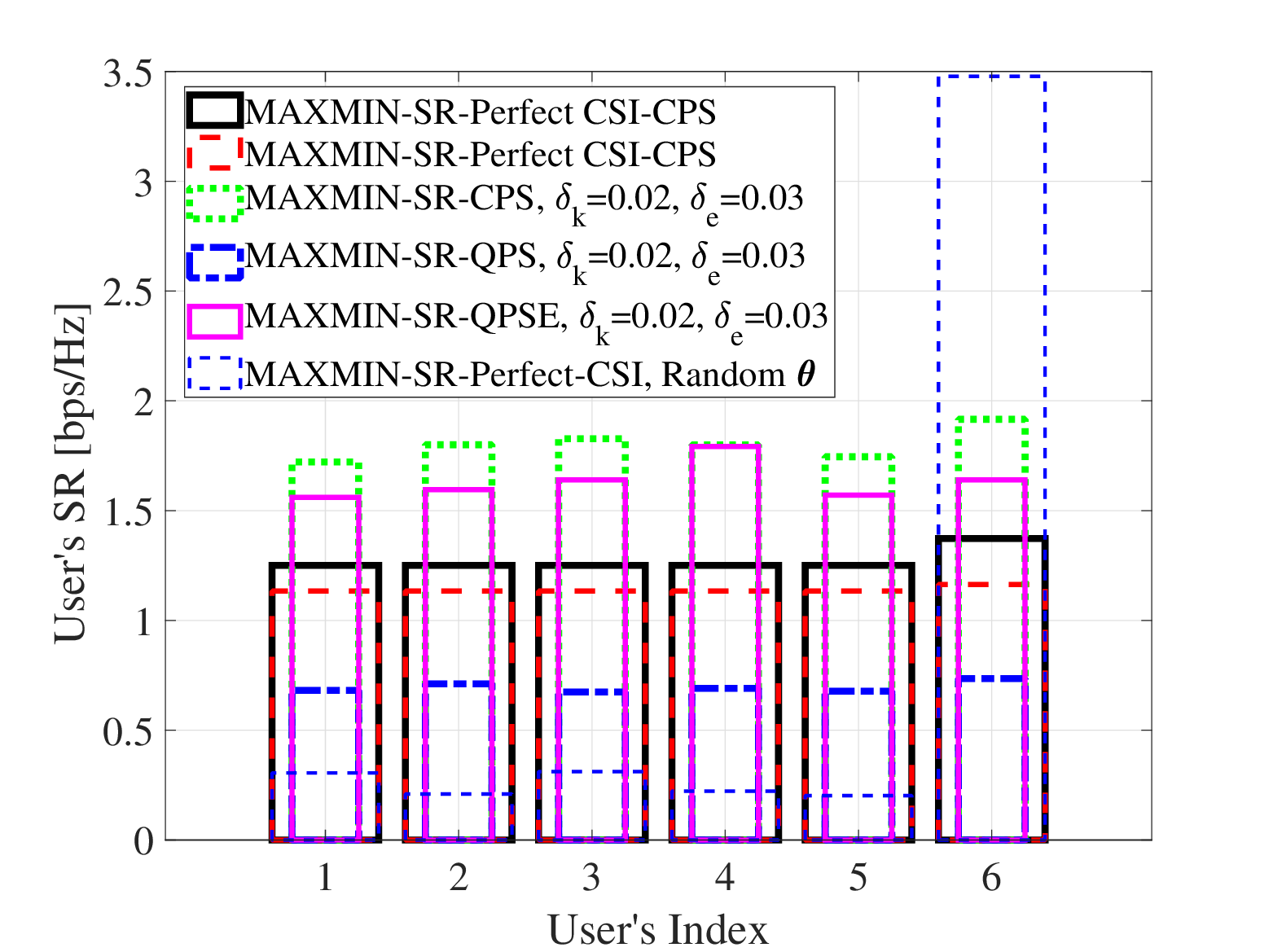}
		\caption{Users' SR with $M=10$, $\mathcal{K}=6$, $N=16$.}
		\label{Fig4}
	\end{figure}

\begin{figure}[t!] 
		\centering
		\includegraphics[width=0.33\textwidth]{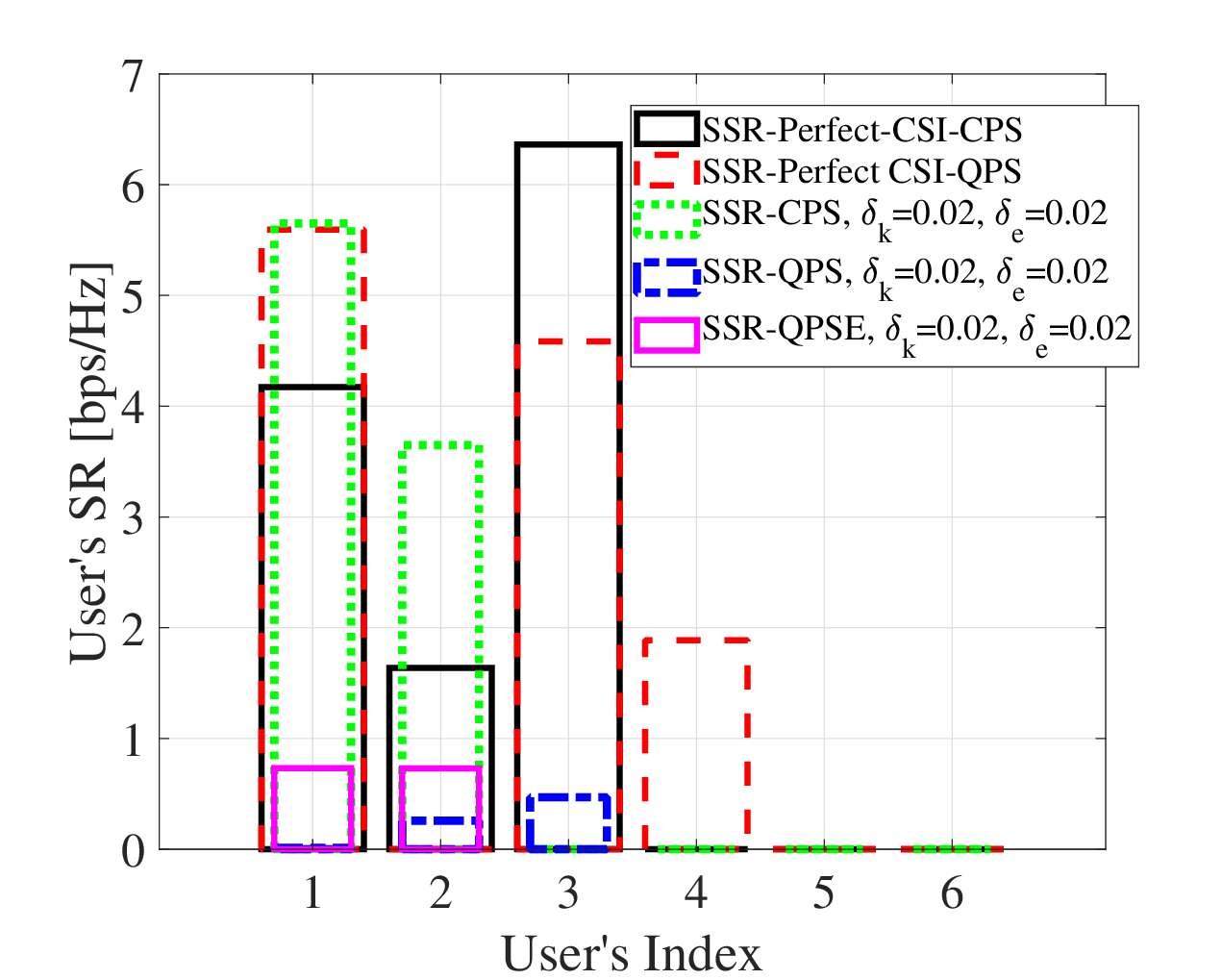}
		\caption{Users' SR with $M=10$, $\mathcal{K}=7$, $N=16$.}
		\label{FigSSRusers}
	\end{figure}

	Fig. \ref{Fig4} shows the users' SR distribution with $M=10$, $\mathcal{K}=6$, and $N=16$, under the Max-Min algorithms for perfect and imperfect CSI cases with CPS, QPS, and QPSE. The CPS achieves the highest fairness between the three cases, which is expected. However, both the QPS and QPSE cases achieved secrecy for \emph{all users} with less fairness compared to the CPS case, and the QPSE case, which has the least fairness in SR among the three cases. However, when the IRS's PREs are not practically optimized, the algorithms fail to achieve secrecy fairness among the users. The results demonstrate the importance of properly optimizing the IRS's PREs to achieve secrecy fairness. The users' SR under the SSR maximization counterparts is shown in Fig. \ref{FigSSRusers}. The SSR maximization algorithms fail to provide secrecy to \emph{all users} and focus most of the transmission power toward the users with better channels.
	
\begin{figure}[t!] 
		\centering
		\includegraphics[width=0.33\textwidth]{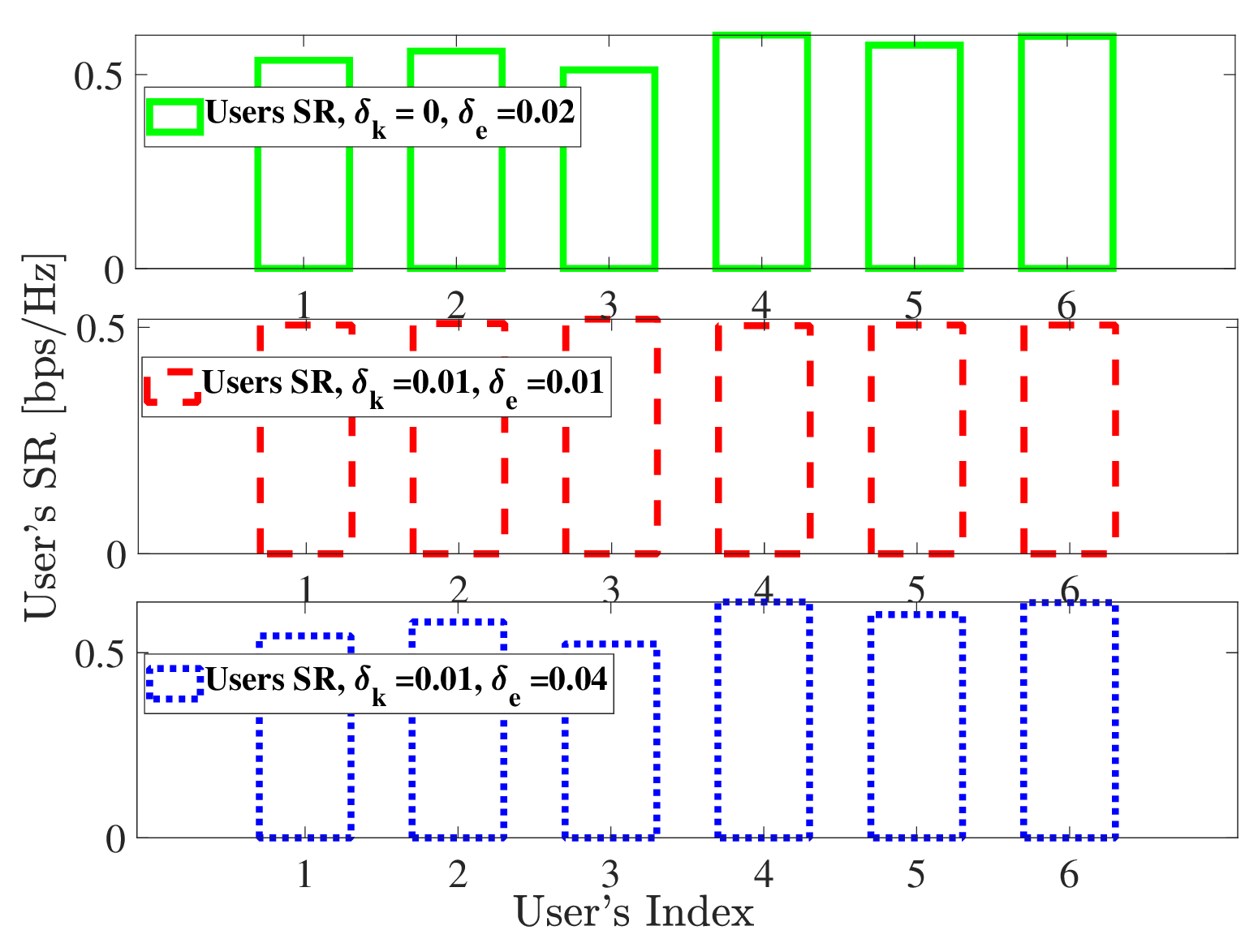}
		\caption{Users' SR with with $M=10, \mathcal{K}=6, N=16$.}
		\label{Fig8}
	\end{figure}

	Fig. \ref{Fig8} depicts the users' SR under various values of the IRS to the users/Eve's imperfect CSI. The Max-Min algorithm achieves secrecy fairness in all cases, even when Eve's reflected channel has higher uncertainty than the users' reflected channel. 
	
\begin{figure}[t!] 
		\centering
		\includegraphics[width=0.33\textwidth]{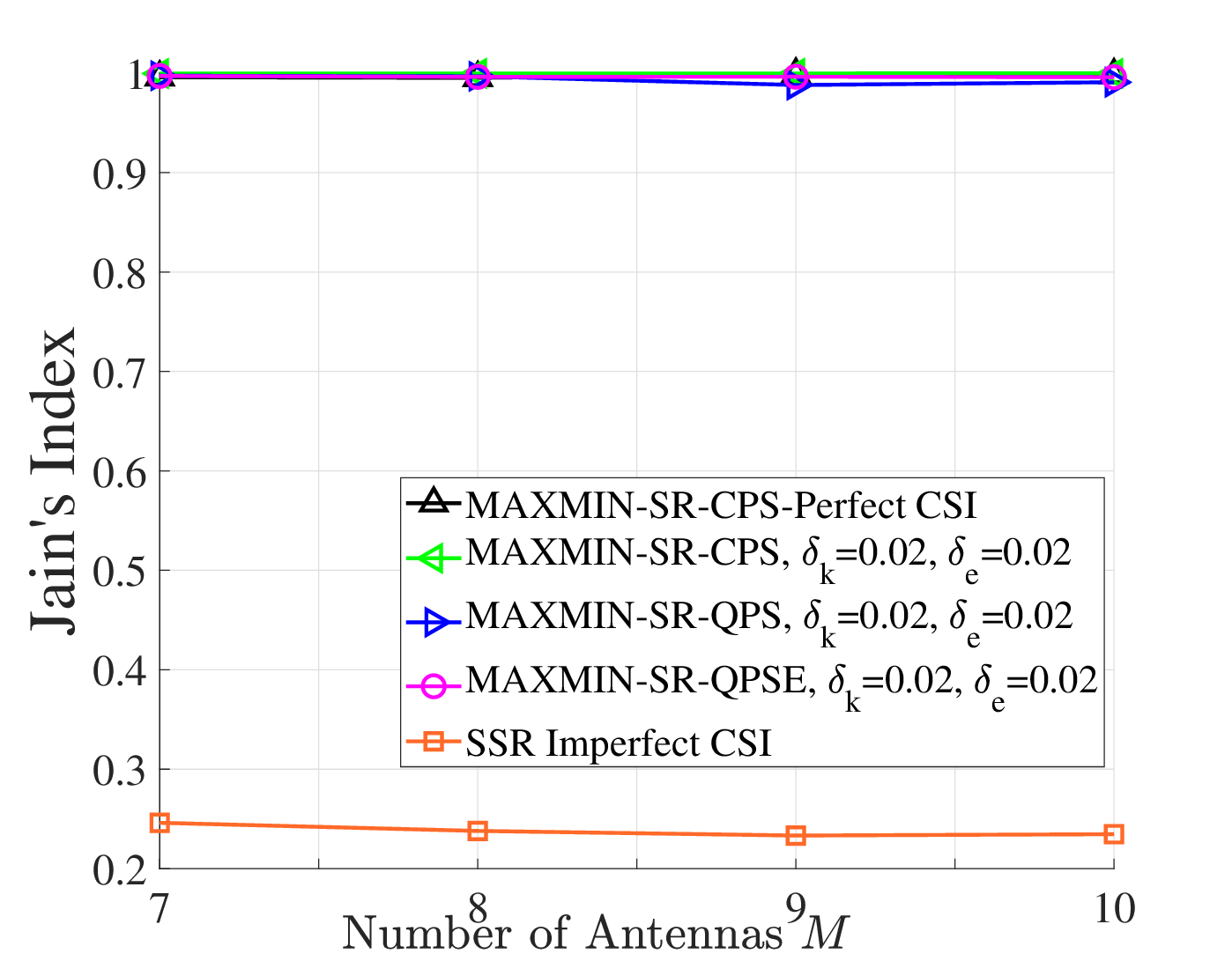}
		\caption{Jain's Index Vs the number of Alice's antennas $M$, with $ \mathcal{K}=4, N=16$.}
		\label{Fig8x}
	\end{figure}

	To evaluate the degree of fairness in the proposed algorithms under imperfect CSI using the Jain's index. Jain's index is defined as, $\text{Jain's Index}= \frac{\left(\sum_{k=1}^{\mathcal{K}}\mbox{SR}_{k}(\pmw,\btheta)\right)^2}{\mathcal{K} \sum_{k=1}^{\mathcal{K}}\left(\mbox{SR}_{k}(\pmw,\btheta)\right)^2}$, and it is bounded in $[1/\mathcal{K},1]$, where the higher value indicates a better fairness \cite{jain1984quantitative}. Fig. \ref{Fig8x} depicts Jain's index while varying the number of Alice's antennas $M$, with $M=10$, $\mathcal{K}=6$, and $N=16$. The proposed Max-Min algorithm achieves almost one for the Jain's index in the CPS/QPS/QPSE cases. As expected, the SSR counterpart achieves low fairness between the users since it favors most of the transmission power $P_T$ towards the users with a better channel while discarding other users, which matches the previous results.
	\begin{figure}[t!] 
		\centering
		\includegraphics[width=0.33\textwidth]{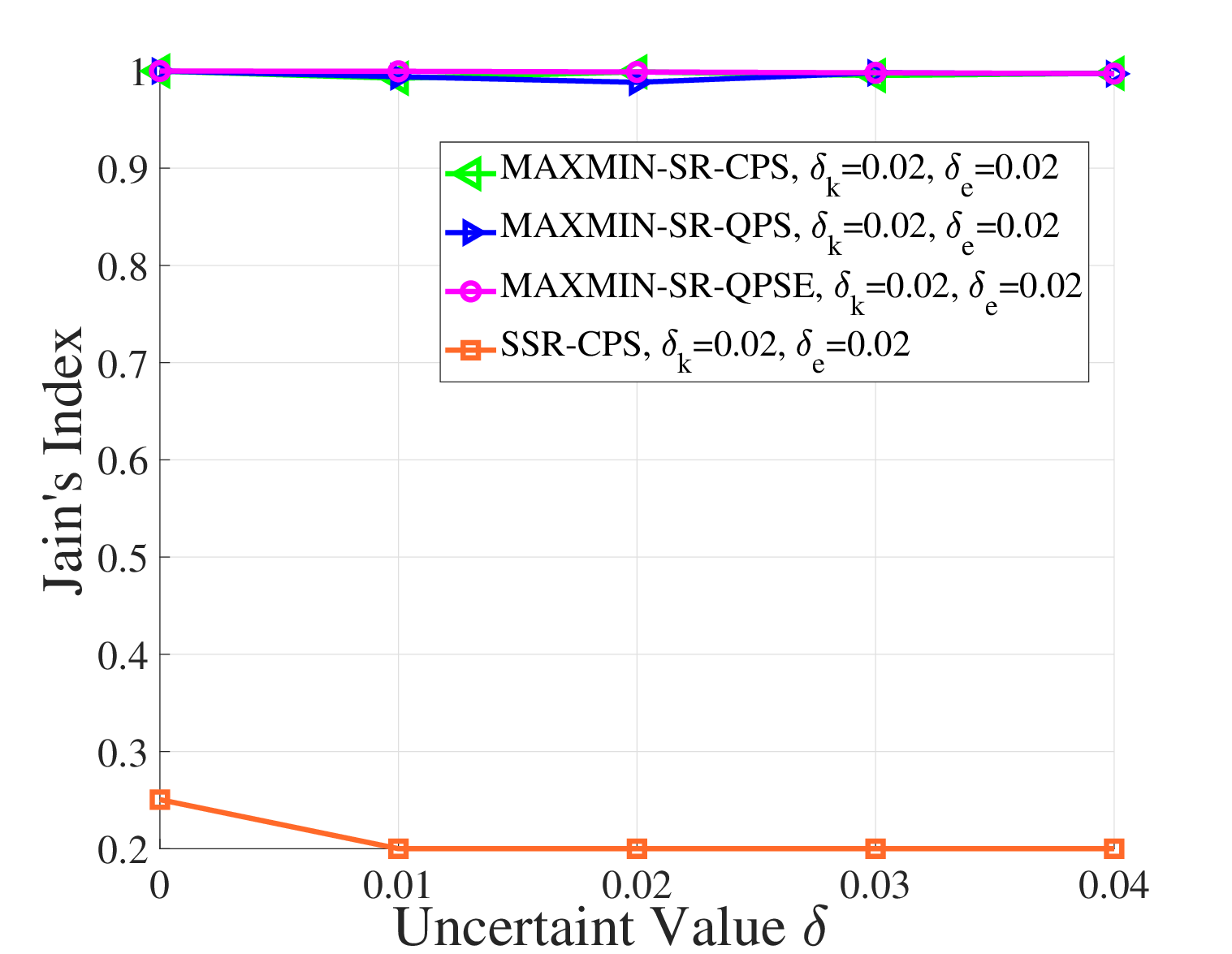}
		\caption{Jain's Index Vs. the uncertainty level $\delta_k=\delta_e$ with $M=10,\mathcal{K}=5, N=16$.}
		\label{Fig9}
	\end{figure}
		\begin{figure*}[!t] 
		\normalsize 
		\begin{align}
			&\ln\left|I_n+[\pmb{A}]^2(\pmb{B})^{-1}\right|\geq \ln\left|I_n+[\hat{A}]^2(\hat{B})^{-1}\right|
			-\la [\hat{A}]^2(\hat{B})^{-1}\ra+2\Re\{\la \hat{A}^H(\hat{B})^{-1}\mathbf{A}\ra\}
			-\la (\hat{B})^{-1}-(\hat{B}+[\hat{A}]^2)^{-1},[\mathbf{A}]^2+\pmb{B}\ra , \label{fund1}  \\ 
			&\ln(1+{\sum}_{i=1}^{l} |{a}_i|^2) \geq \ln(1+{\sum}_{i=1}^{l} |\bar{a}_i|^2)-{\sum}_{i=1}^{l} |\bar{a}_i|^2 +{\sum}_{i=1}^{l} 2 \Re\{\bar{a}_i^* {a}_i\} - \frac{{\sum}_{i=1}^{l} |\bar{a}_i|^2 \left(1+{\sum}_{i=1}^{l} |{a}_i|^2\right)}{1+{\sum}_{i=1}^{l} |\bar{a}_i|^2}.  \label{fund2} 
		\end{align} 
		\hrulefill 
	\end{figure*}
	
	Fig. \ref{Fig9} shows the Jain's index against the relative amount of CSI uncertainty $\delta$. The Max-Min algorithm achieves almost one in all cases, CPS/QPS/QPSE, while the SSR counterpart has nearly zero. The results demonstrate the Max-Min algorithm's robustness even when the phase error is introduced at high uncertainty CSI for the users and Eve counterpart. 
	
\begin{figure}[t!] 
		\centering
		\includegraphics[width=0.33\textwidth]{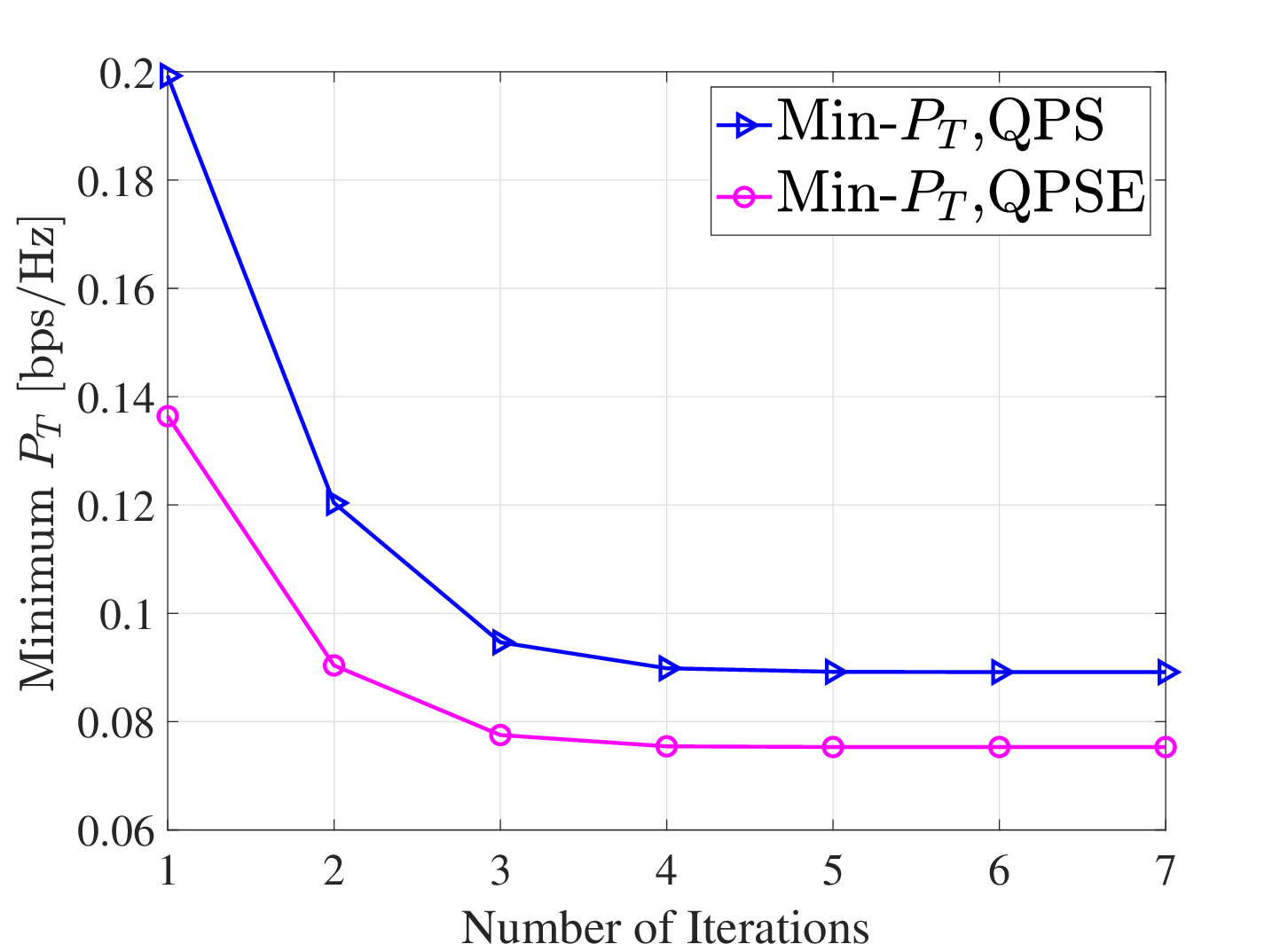}
		\caption{Convergence rate of the Min $P_T$ algorithm with $M=10$, $\mathcal{K}=5$, $N=16$, $b=3$, $\delta_k=0.01$,$\gamma_k=0.5[\text{dB}]$.}
		\label{Fig12}
	\end{figure}

\begin{figure}[t!] 
		\centering
		\includegraphics[width=0.33\textwidth]{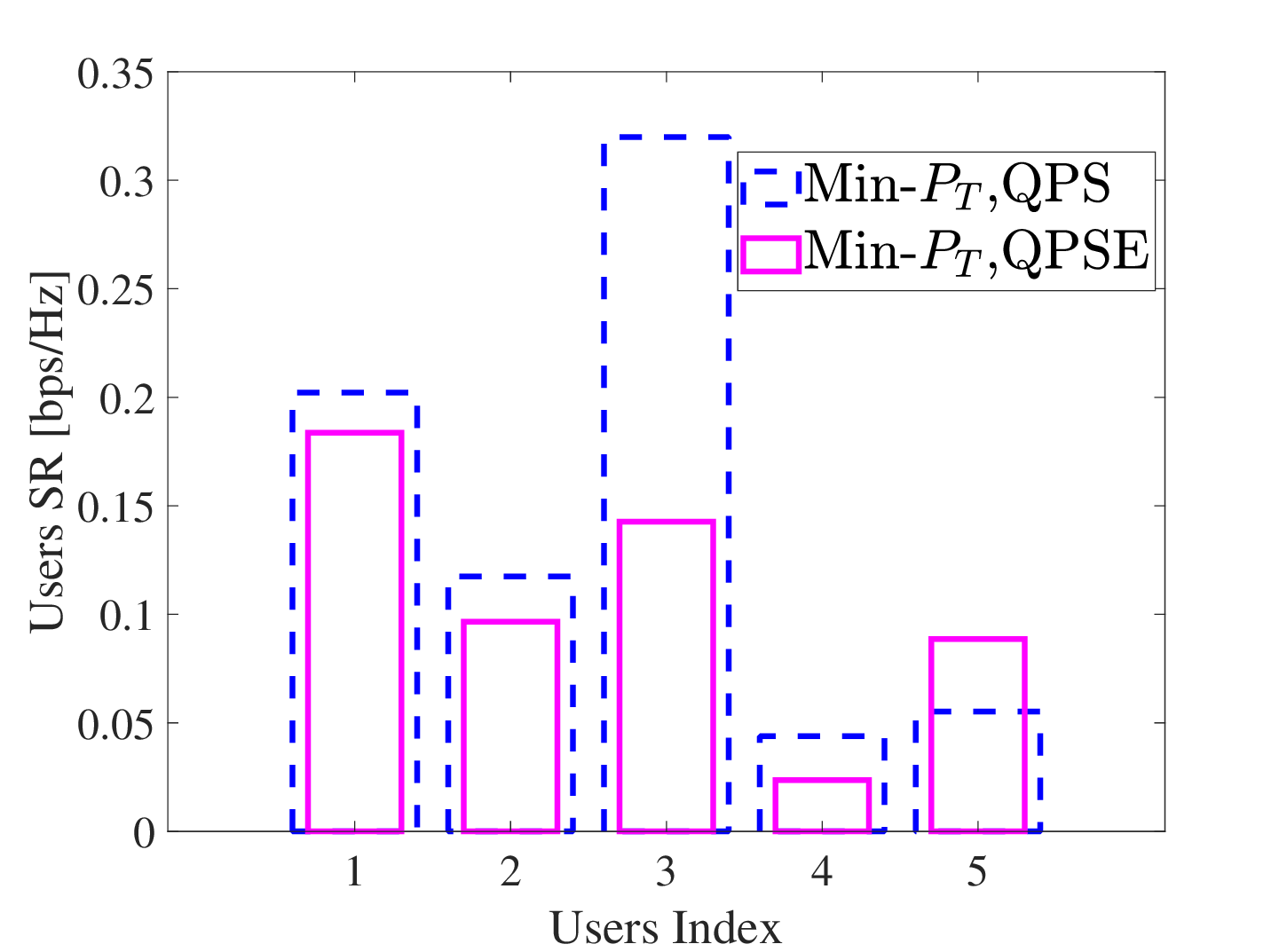}
		\caption{Users' SR with with $M=10$, $\mathcal{K}=5$, $N=16$, $b=3$, $\delta_k=0.01$, $\gamma_k=0.5 [dB]$.}
		\label{Fig10}
	\end{figure}

	Fig. \ref{Fig12} illustrates the convergence rate of Algorithm \ref{alg7} with the QPS and the QPSE cases. The algorithm converges within a few iterations.  
	Fig. \ref{Fig10} shows the users' SR distribution with $M=10$, $\mathcal{K}=5$, and $N=16$, under the power minimization algorithm for the imperfect CSI case with  QPS, and QPSE. \emph{All users} have achieved secrecy while maintaining the minimum QoS constraint $\gamma_k=0.5[\text{dB}]$, which demonstrates the robustness of the proposed algorithm even with QPSE and the users' imperfect CSI introduced, while Eve's CSI is unknown to Alice.
	\begin{figure}[t!] 
		\centering
		\includegraphics[width=0.33\textwidth]{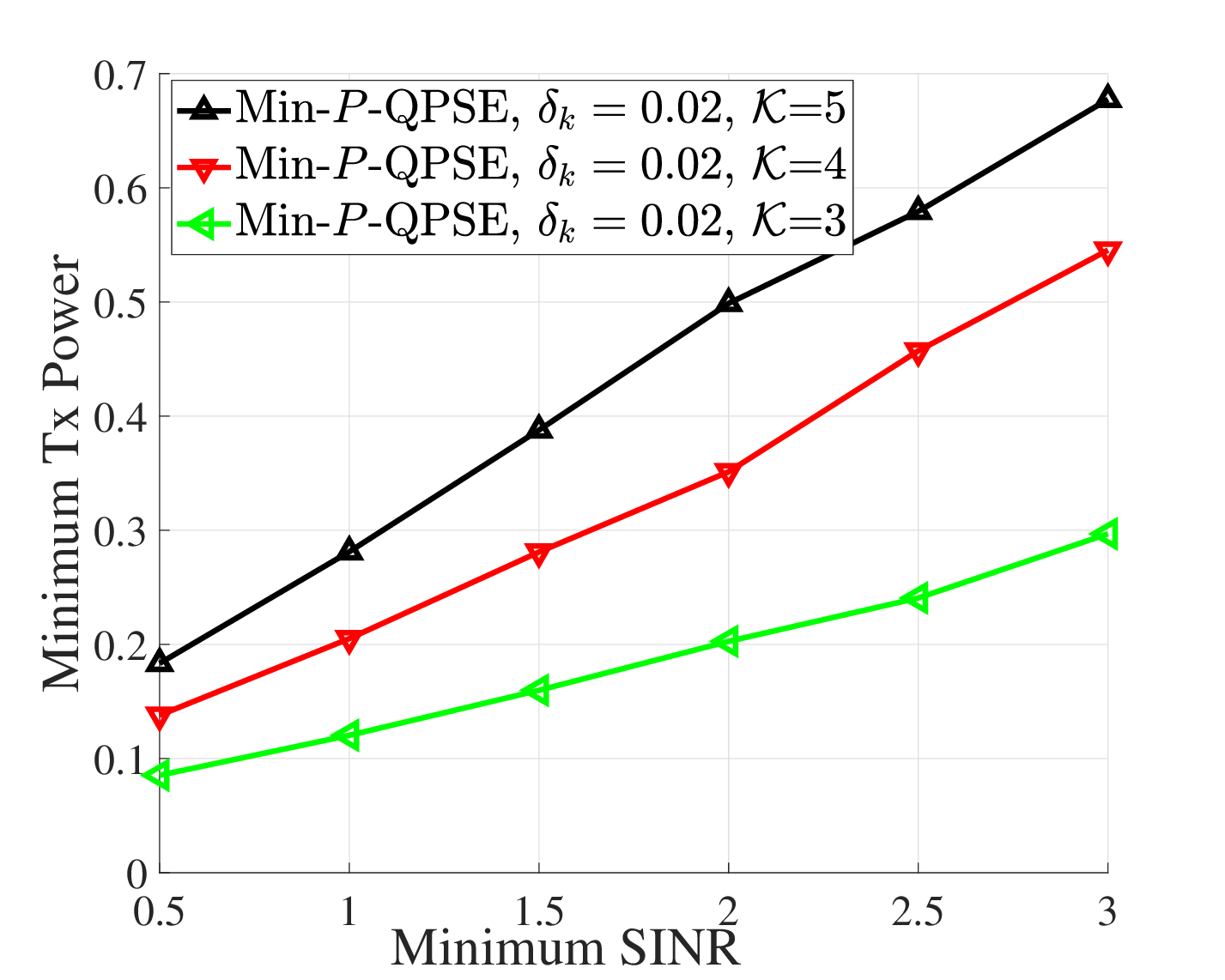}
		\caption{Min-Tx-Power Vs. the minimum users' SINR $\gamma_k$ with $M=10$, $N=25$, $b=3$.}
		\label{Fig11}
	\end{figure}
	
	Lastly, Fig. \ref{Fig11} depicts the minimum transmit power $P_T$ against the minimum user's SINR $\gamma_k$. The minimum power increases with $\gamma_k$ as expected, since the system requires more transmit power to achieve higher users' SINR.  The results show that the system will require higher transmit power to achieve secrecy for \emph{all users} with increasing $\mathcal{K}$, and it can achieve secrecy even when the PREs are subject to the QPSE.
	\vspace{-0.5cm}\section{Conclusions} \label{conc}
	In this paper, we proposed a framework to achieve secure communications for all the users in a low-resolution IRS-aided system with the presence of an eavesdropper under both perfect and imperfect CSI. Specifically, through linearization and different non-convex optimization techniques, we designed computationally efficient algorithms to maximize the minimum SR among \emph{all users} under both perfect and imperfect CSI from IRS to the users and the eavesdropper, and minimize the transmit power subject to minimum SR constraints, where the IRS's PREs are modeled by quantized phase shift (QPS) and affected by phase shift error (PSE).  Extensive simulation results showed that the Max-Min algorithm can provide secure communications for all the users even with only imperfect CSI. In the future, one can consider a multi-hop scenario with the joint coding over multiple IRSs or distributed beamforming, similar to the cooperative MIMO setting \cite{vmimo}.
	\begin{appendices}
		\section{Proof of theorem 1} \label{AppB}
		First we prove that the sequence $\mbox{SR}_k(\wko,\lfloor \thetako \rceil_b+\epsilon_R)$ is non decreasing for all $k$, i.e.  $\mbox{SR}_k(\wko,\lfloor \thetako \rceil_b) \geq \mbox{SR}_k(\wk,\lfloor \thetak \rceil_b)$ for all $\iota >0$. To establish this, $\wko$ is obtained by solving problem ($\mathcal{P}1.2$) via the CVX solver, which guarantees finding the optimal solution. Thus, we have $\mbox{SR}_k(\wko,\lfloor\btheta\rceil_b+\epsilon_R) > \mbox{SR}_k(\wk,\btheta)$ for all $k$. Next, by solving problem ($\mathcal{P}1.3$), we update the IRS PREs' phase shift to $\lfloor \thetako_n \rceil_b+\epsilon_R$, yielding $\mbox{SR}_k(\pmw,\lfloor \thetako_n \rceil_b+\epsilon_R) > \mbox{SR}_k(\pmw,\lfloor\thetak\rceil_b+\epsilon_R)$ as stated before. 
		
		Hence, by combining the solutions, we obtain 
		\begin{align*} 
			\mbox{SR}_k(\wko,\lfloor \thetako_n \rceil_b+\epsilon_R) > \mbox{SR}_k(\wk,\lfloor\thetak\rceil_b+\epsilon_R),
		\end{align*}
		hence, the optimal sequence $\{(\wko,\lfloor \thetako_n \rceil_b+\epsilon_R)\}$ converge to a point $\{({\pmw}^*,\lfloor{\btheta}^*\rceil_b+\epsilon_R)\}$ which is the solution obtained from solving ($\mathcal{P}2$) and ($\mathcal{P}1.1$). 
		
		Next, we prove that the converged point $\hat{X}^* \triangleq \{\pmw^*,\btheta^*\}$ is a locally optimal solution of problem $\mathcal{P}1$. For that, we show that the converged point satisfies the Karush-Kuhn-Tucker (KKT) condition of the problem. In particular, the KKT conditions for problem $(\mathcal{P}1.1)$ hold at $\btheta^*$. Let $\pmb{Y}(\btheta)$ is the objective function of $(\mathcal{P}3)$ and $\pmb{T}({\pmb{X}})\triangleq[\pmb{T}_1({\pmb{X}}), \pmb{T}_2({\pmb{X}}), \dots, \pmb{T}_I({\pmb{X}})]$ be the set of constraints of problem $\mathcal{P}3$. Then, we can write, 
		\begin{align} \label{con1}
			&\nabla_{\btheta^*}\pmb{Y}({\pmb{X}}^*)+ \pmb{Z}^T \nabla_{\btheta^*}\pmb{T}({\pmb{X}}^*) =0, \\
			& z_i \geq 0, z_i \pmb{T}({\pmb{X}}^*)=0, \forall i. \nonumber
		\end{align}
		where $\pmb{Z} \triangleq [z_1,z_2,\dots,z_I]$ is the optimal Lagrangian variable set, and $\nabla_s$ is the gradient with respect to $s$. Similarly, the solution obtained by solving $(\mathcal{P}2)$ is locally optimal. Hence its KKT is satisfied with respect to $\pmb{\text{W}}=\pmw_k^*$, which is, 
		\begin{align} \label{con2}
			&\nabla_{\pmb{\text{W}}^*}\pmb{Y}({\pmb{X}}^*)+ Z^T \nabla_{\pmb{\text{W}}^*}\pmb{T}({\pmb{X}}^*) =0, \\
			& z_i \geq 0, t_i \pmb{T}({\pmb{X}}^*)=0, \forall i. \nonumber
		\end{align}
		Combining \eqref{con1} and \eqref{con2}, we get
		\begin{align*}
			&\nabla_{{\pmb{X}}^*}\pmb{Y}({\pmb{X}}^*)+ Z^T \nabla_{{\pmb{X}}^*}\pmb{T}({\pmb{X}}^*) =0, \\
			& t_i \geq 0, t_i \pmb{T}({\pmb{X}}^*)=0, \forall i, \nonumber
		\end{align*}
		which is the KKT condition for $(\mathcal{P}1)$. $\blacksquare$
		
		\section{Proof of Lemma 1} \label{apendixB}
		Let $x$ be a scalar complex variable, and denote by $x^{(\iota)}$ the fixed point obtained at iteration $(\iota)$. The following inequality holds \cite{9110587},
		\vspace{-0.2cm}\begin{align}
			|x|^2 \geq x^{*,(\iota)} x +x^{*} x^{(\iota)} - x^{*,(\iota)} x^{(\iota)},
		\end{align}
		By replacing $x$ with $({{\pmb{g}}}_k \pmb{\Phi} {\pmb{H}_\text{AR}})\pmw_k$ we obtain \eqref{Bobx1}. Thus, this concludes the proof. $\blacksquare$
		\vspace{-0.2cm}\section{Proof of theorem 2, and 3} \label{AppD}
		Similar to the proof of Theorem \ref{theo1}, it can be shown the sequence $\mbox{SR}_k(\wko,\lfloor \thetako_n \rceil_b)$ is non-decreasing for all $k$, i.e.  $\mbox{SR}_k(\wko,\lfloor \thetako_n \rceil_b) \geq \mbox{SR}_k(\wk,\lfloor\thetak\rceil_{b})$ for all $(\iota) >0$, and the converged point is a locally optimal solution for problems. $\blacksquare$.
		\vspace{-0.4cm}\section{Inequalities} \label{AppA}
		Here, we utilize the inequality (48) in \cite{TTN16}. Specifically, for any $\mathbf{X}$ and $\pmb{Y}$ with $\hat{A}$ and $\hat{B}$ are fixed point, \eqref{fund1} holds.
		
		Next, we invoke Lemma (2) in \cite{niu2022joint}. Specifically, for any ${a}_b$ with $i=1,\dots,l$ and $\bar{a}_b$ is a fixed point, inequality \eqref{fund2} holds.
		Lastly, the logarithmic function is a concave function and can be written as \cite[Eq. 15]{niu2022joint}
		\begin{align} \label{fund3}
			-\ln(1+\pmb{{\varsigma}}) \geq -\ln(1+\bar{\varsigma}) - {1+\pmb{\varsigma}}/{1+\bar{\varsigma}} +1.
		\end{align}

	\end{appendices}
	
	\bibliographystyle{IEEEtran}
	\bibliography{surface}

\begin{thebibliography}{10}
\providecommand{\url}[1]{#1}
\csname url@samestyle\endcsname
\providecommand{\newblock}{\relax}
\providecommand{\bibinfo}[2]{#2}
\providecommand{\BIBentrySTDinterwordspacing}{\spaceskip=0pt\relax}
\providecommand{\BIBentryALTinterwordstretchfactor}{4}
\providecommand{\BIBentryALTinterwordspacing}{\spaceskip=\fontdimen2\font plus
\BIBentryALTinterwordstretchfactor\fontdimen3\font minus \fontdimen4\font\relax}
\providecommand{\BIBforeignlanguage}[2]{{%
\expandafter\ifx\csname l@#1\endcsname\relax
\typeout{** WARNING: IEEEtran.bst: No hyphenation pattern has been}%
\typeout{** loaded for the language `#1'. Using the pattern for}%
\typeout{** the default language instead.}%
\else
\language=\csname l@#1\endcsname
\fi
#2}}
\providecommand{\BIBdecl}{\relax}
\BIBdecl

\bibitem{10901599}
M.~Abughalwa, D.~N. Nguyen, D.~T. Hoang, and E.~Dutkiewicz, ``Multi-user secrecy rate maximization in {IRS}-aided systems,'' in \emph{GLOBECOM 2024 - 2024 IEEE Global Communications Conference}, 2024, pp. 2491--2496.

\bibitem{nguyen2022leveraging}
T.~V. Nguyen, D.~N. Nguyen, M.~D. Renzo, and R.~Zhang, ``Leveraging secondary reflections and mitigating interference in multi-{IRS}/{RIS} aided wireless networks,'' \emph{IEEE Transactions on Wireless Communications}, vol.~22, no.~1, pp. 502--517, 2023.

\bibitem{abughalwa2022finite}
M.~Abughalwa, H.~D. Tuan, D.~N. Nguyen, H.~V. Poor, and L.~Hanzo, ``Finite-blocklength {RIS}-aided transmit beamforming,'' \emph{IEEE Transactions on Vehicular Technology}, vol.~71, no.~11, pp. 12\,374--12\,379, 2022.

\bibitem{9896755}
G.~Chen, Q.~Wu, C.~He, W.~Chen, J.~Tang, and S.~Jin, ``Active {IRS} aided multiple access for energy-constrained {IoT} systems,'' \emph{IEEE Transactions on Wireless Communications}, vol.~22, no.~3, pp. 1677--1694, 2023.

\bibitem{zhou2021secure}
G.~Zhou, C.~Pan, H.~Ren, K.~Wang, and Z.~Peng, ``Secure wireless communication in {RIS}-aided {MISO} system with hardware impairments,'' \emph{IEEE Wireless Communications Letters}, vol.~10, no.~6, pp. 1309--1313, 2021.

\bibitem{hong2020robust}
S.~Hong, C.~Pan, H.~Ren, K.~Wang, K.~K. Chai, and A.~Nallanathan, ``Robust transmission design for intelligent reflecting surface-aided secure communication systems with imperfect cascaded {CSI},'' \emph{IEEE Transactions on Wireless Communications}, vol.~20, no.~4, pp. 2487--2501, 2020.

\bibitem{9913501}
W.~Shi, Q.~Wu, F.~Xiao, F.~Shu, and J.~Wang, ``Secrecy throughput maximization for {IRS}-aided {MIMO} wireless powered communication networks,'' \emph{IEEE Transactions on Communications}, vol.~70, no.~11, pp. 7520--7535, 2022.

\bibitem{10256584}
W.~Li, W.~Yu, H.~Liu, and H.~Hou, ``Robust secrecy rate maximization for {IRS}-aided {MISO} communication systems,'' in \emph{2023 IEEE 13th International Conference on CYBER Technology in Automation, Control, and Intelligent Systems (CYBER)}, 2023, pp. 604--609.

\bibitem{9402750}
L.~Dong, H.-M. Wang, and H.~Xiao, ``Secure cognitive radio communication via intelligent reflecting surface,'' \emph{IEEE Transactions on Communications}, vol.~69, no.~7, pp. 4678--4690, 2021.

\bibitem{9180053}
G.~Zhou, C.~Pan, H.~Ren, K.~Wang, and A.~Nallanathan, ``A framework of robust transmission design for {IRS}-aided {MISO} communications with imperfect cascaded channels,'' \emph{IEEE Transactions on Signal Processing}, vol.~68, pp. 5092--5106, 2020.

\bibitem{7510962}
X.~Tan, Z.~Sun, J.~M. Jornet, and D.~Pados, ``Increasing indoor spectrum sharing capacity using smart reflect-array,'' in \emph{2016 IEEE International Conference on Communications (ICC)}, 2016, pp. 1--6.

\bibitem{8683145}
Q.~Wu and R.~Zhang, ``Beamforming optimization for intelligent reflecting surface with discrete phase shifts,'' in \emph{ICASSP 2019 - 2019 IEEE International Conference on Acoustics, Speech and Signal Processing ()}, 2019, pp. 7830--7833.

\bibitem{8746155}
Y.~Han, W.~Tang, S.~Jin, C.-K. Wen, and X.~Ma, ``Large intelligent surface-assisted wireless communication exploiting statistical {CSI},'' \emph{IEEE Transactions on Vehicular Technology}, vol.~68, no.~8, pp. 8238--8242, 2019.

\bibitem{8869792}
M.-A. Badiu and J.~P. Coon, ``Communication through a large reflecting surface with phase errors,'' \emph{IEEE Wireless Communications Letters}, vol.~9, no.~2, pp. 184--188, 2020.

\bibitem{9197675}
F.~Fang, Y.~Xu, Q.-V. Pham, and Z.~Ding, ``Energy-efficient design of {IRS}-{NOMA} networks,'' \emph{IEEE Transactions on Vehicular Technology}, vol.~69, no.~11, pp. 14\,088--14\,092, 2020.

\bibitem{boyd1994linear}
S.~Boyd, L.~El~Ghaoui, E.~Feron, and V.~Balakrishnan, \emph{Linear matrix inequalities in system and control theory}.\hskip 1em plus 0.5em minus 0.4em\relax SIAM, 1994.

\bibitem{9525400}
H.~Niu, Z.~Chu, F.~Zhou, and Z.~Zhu, ``Simultaneous transmission and reflection reconfigurable intelligent surface assisted secrecy {MISO} networks,'' \emph{IEEE Communications Letters}, vol.~25, no.~11, pp. 3498--3502, 2021.

\bibitem{9110587}
G.~Zhou, C.~Pan, H.~Ren, K.~Wang, M.~D. Renzo, and A.~Nallanathan, ``Robust beamforming design for intelligent reflecting surface aided {MISO} communication systems,'' \emph{IEEE Wireless Communications Letters}, vol.~9, no.~10, pp. 1658--1662, 2020.

\bibitem{kammoun2020asymptotic}
Q.-U.-A. Nadeem, A.~Kammoun, A.~Chaaban, M.~Debbah, and M.-S. Alouini, ``Asymptotic max-min {SINR} analysis of reconfigurable intelligent surface assisted {MISO} systems,'' \emph{IEEE Transactions on Wireless Communications}, vol.~19, no.~12, pp. 7748--7764, 2020.

\bibitem{9501057}
B.~Zheng, C.~You, and R.~Zhang, ``Uplink channel estimation for double-irs assisted multi-user mimo,'' in \emph{ICC 2021 - IEEE International Conference on Communications}, 2021, pp. 1--6.

\bibitem{9373363}
------, ``Efficient channel estimation for double-irs aided multi-user mimo system,'' \emph{IEEE Transactions on Communications}, vol.~69, no.~6, pp. 3818--3832, 2021.

\bibitem{8579566}
C.~Pan, H.~Ren, M.~Elkashlan, A.~Nallanathan, and L.~Hanzo, ``Robust beamforming design for ultra-dense user-centric {C-RAN} in the face of realistic pilot contamination and limited feedback,'' \emph{IEEE Transactions on Wireless Communications}, vol.~18, no.~2, pp. 780--795, 2019.

\bibitem{9266086}
Z.~Zhang, L.~Lv, Q.~Wu, H.~Deng, and J.~Chen, ``Robust and secure communications in intelligent reflecting surface assisted {NOMA} networks,'' \emph{IEEE Communications Letters}, vol.~25, no.~3, pp. 739--743, 2021.

\bibitem{bloch2008wireless}
M.~Bloch, J.~Barros, M.~R.~D. Rodrigues, and S.~W. McLaughlin, ``Wireless information-theoretic security,'' \emph{IEEE Transactions on Information Theory}, vol.~54, no.~6, pp. 2515--2534, 2008.

\bibitem{niu2022joint}
H.~Niu, Z.~Lin, Z.~Chu, Z.~Zhu, P.~Xiao, H.~X. Nguyen, I.~Lee, and N.~Al-Dhahir, ``Joint beamforming design for secure {RIS}-assisted {IoT} networks,'' \emph{IEEE Internet of Things Journal}, vol.~10, no.~2, pp. 1628--1641, 2023.

\bibitem{grant2014cvx}
M.~Grant and S.~Boyd, ``Cvx: Matlab software for disciplined convex programming, version 2.1,'' 2014.

\bibitem{labit2002sedumi}
Y.~Labit, D.~Peaucelle, and D.~Henrion, ``Sedumi interface 1.02: a tool for solving {LMI} problems with sedumi,'' in \emph{Proceedings. IEEE International Symposium on Computer Aided Control System Design}.\hskip 1em plus 0.5em minus 0.4em\relax IEEE, 2002, pp. 272--277.

\bibitem{TTN16}
H.~H.~M. Tam, H.~D. Tuan, and D.~T. Ngo, ``Successive convex quadratic programming for quality-of-service management in full-duplex {MU}-{MIMO} multicell networks,'' \emph{IEEE Transactions on Communications}, vol.~64, no.~6, pp. 2340--2353, 2016.

\bibitem{boyd2004convex}
S.~Boyd and L.~Vandenberghe, \emph{Convex optimization}.\hskip 1em plus 0.5em minus 0.4em\relax Cambridge university press, 2004.

\bibitem{1369660}
Y.~Eldar, A.~Ben-Tal, and A.~Nemirovski, ``Robust mean-squared error estimation in the presence of model uncertainties,'' \emph{IEEE Transactions on Signal Processing}, vol.~53, no.~1, pp. 168--181, 2005.

\bibitem{9774882}
W.~Wang, W.~Ni, H.~Tian, Z.~Yang, C.~Huang, and K.-K. Wong, ``Safeguarding {NOMA} networks via reconfigurable dual-functional surface under imperfect {CSI},'' \emph{IEEE Journal of Selected Topics in Signal Processing}, vol.~16, no.~5, pp. 950--966, 2022.

\bibitem{9505311}
Y.~Chen, Y.~Wang, and L.~Jiao, ``Robust transmission for reconfigurable intelligent surface aided millimeter wave vehicular communications with statistical {CSI},'' \emph{IEEE Transactions on Wireless Communications}, vol.~21, no.~2, pp. 928--944, 2022.

\bibitem{9146177}
H.-M. Wang, J.~Bai, and L.~Dong, ``Intelligent reflecting surfaces assisted secure transmission without eavesdropper's csi,'' \emph{IEEE Signal Processing Letters}, vol.~27, pp. 1300--1304, 2020.

\bibitem{9764813}
H.~Niu, X.~Lei, Y.~Xiao, M.~Xiao, and S.~Mumtaz, ``On the efficient design of {RIS}-assisted secure {MISO} transmission,'' \emph{IEEE Wireless Communications Letters}, vol.~11, no.~8, pp. 1664--1668, 2022.

\bibitem{6772207}
A.~D. Wyner, ``The wire-tap channel,'' \emph{The Bell System Technical Journal}, vol.~54, no.~8, pp. 1355--1387, 1975.

\bibitem{BOL19}
E.~Bj{\H o}rnson, {\H O}.~{\H O}zdogan, and E.~G. Larsson, ``Intelligent reflecting surface versus decode-and-forward: How large surfaces are needed to beat relaying?'' \emph{IEEE Wireless Communications Letters}, vol.~9, no.~2, pp. 244--248, 2020.

\bibitem{jain1984quantitative}
R.~K. Jain, D.-M.~W. Chiu, W.~R. Hawe \emph{et~al.}, ``A quantitative measure of fairness and discrimination,'' \emph{Eastern Research Laboratory, Digital Equipment Corporation, Hudson, MA}, vol.~21, 1984.

\bibitem{vmimo}
\BIBentryALTinterwordspacing
D.~N. Nguyen and M.~Krunz, ``A cooperative mimo framework for wireless sensor networks,'' \emph{ACM Trans. Sen. Netw.}, vol.~10, no.~3, May 2014. [Online]. Available: \url{https://doi.org/10.1145/2499381}
\BIBentrySTDinterwordspacing

\end{thebibliography}
\end{document}